\documentclass[journal=mamobx,manuscript=article]{achemso}
\setkeys{acs}{articletitle=true}

\SectionNumbersOn

\usepackage[version=3]{mhchem} % Formula subscripts using \ce{}
\usepackage[T1]{fontenc} % Use modern font encodings

\usepackage{color}
\usepackage{gensymb}
\author{Xiaolei Xu}
\affiliation{State Key Laboratory of Polymer Physics and Chemistry, Changchun Institute of Applied Chemistry, Chinese Academy of Sciences, Changchun 130022, P. R. China}

\author{Jack F. Douglas}
\email{jack.douglas@nist.gov}
\affiliation{Materials Science and Engineering Division, National Institute of Standards and Technology, Gaithersburg, Maryland 20899, United States}

\author{Wen-Sheng Xu}
\email{wsxu@ciac.ac.cn}
\affiliation{State Key Laboratory of Polymer Physics and Chemistry, Changchun Institute of Applied Chemistry, Chinese Academy of Sciences, Changchun 130022, P. R. China}
\title{Thermodynamic-Dynamic Interrelations in Glass-Forming Polymer Fluids}

\abbreviations{IR,NMR,UV}

\begin{document}

\begin{tocentry}

 \centering
 \includegraphics[height=3cm]{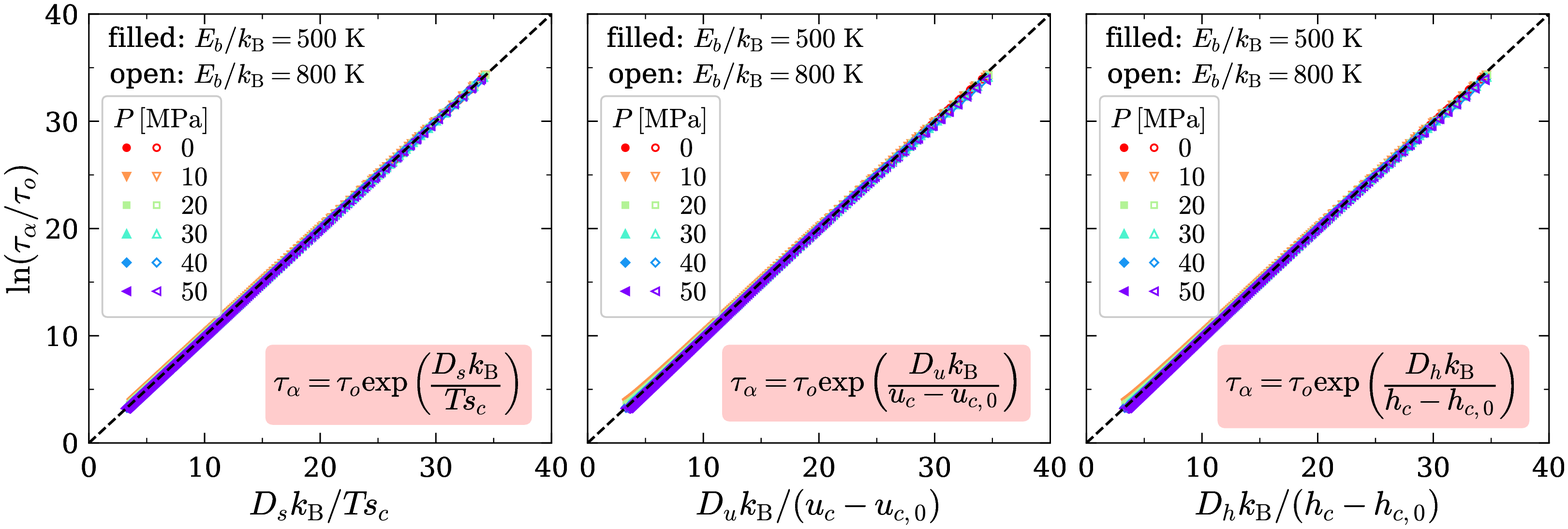}

% Some text to explain the graphic.

\end{tocentry}

\newpage

\begin{abstract}
There is a long history of trying to understand the dynamics of glass-forming and other condensed materials exhibiting highly anharmonic interparticle interactions, based on their thermodynamic properties. This has led to numerous correlations between thermodynamic (e.g., density, compressibility, enthalpy, entropy, and vapor pressure) and dynamic (e.g., viscosity, diffusion coefficients, and relaxation times) properties, and a steady stream of theoretical models has been introduced to rationalize these correlations in the absence of any generally accepted theory of the dynamics of non-crystalline condensed materials. We view the independent success of these various semi-empirical models of glass-forming liquids as possibly pointing to a greater unity arising from the strong interrelation between thermodynamic properties, which is a matter of interest beyond an understanding of the dynamics of glass-forming liquids. Accordingly, we utilize the lattice cluster theory (LCT) of polymer fluids to show that the configurational entropy, enthalpy, and internal energy are all closely interrelated, as suggested by recent measurements by Caruthers and Medvedev, so that the generalized entropy theory (GET) of glass formation, a combination of the LCT and Adam-Gibbs model, can be recast in terms of any of these thermodynamic properties as a matter of convenience. Thermodynamic scaling, a form of density-temperature scaling exhibited by dynamic and some thermodynamic properties, is used to assess which thermodynamic properties are most naturally linked to dynamics, and we explore the origin of this scaling by both direct calculations based on the GET and molecular dynamics simulations of a coarse-grained polymer model. Through a combination of our comprehensive modeling of thermodynamic properties using the LCT and the highly predictive GET model for how the fluid thermodynamics relate to its dynamics, along with simulation results confirming these theoretical frameworks, we obtain insights into thermodynamic aspects of collective motion and the slow $\beta$-relaxation processes of glass-forming liquids.
\end{abstract}

\newpage

\section{\label{Sec_Intro}Introduction}

A basic aim of statistical mechanics is to explain the thermodynamic properties of materials at equilibrium in terms of the structure and interaction parameters of the molecules or particles comprising these materials under specified state conditions, such as temperature ($T$), volume ($V$), pressure ($P$), etc. It also aims at describing transport properties (e.g., viscosity and diffusion) in terms of thermodynamic properties, and significant success in this direction has been achieved in this enterprise in the case of dilute gases through the kinetic theory developed by Maxwell,~\cite{1867_PTRS_157_49} Jeans,~\cite{Book_Jeans} Enskog,~\cite{1970_JCP_53_3813, Book_Chapman, Book_Hirschfelder, Note_GasTheory} etc., and through semi-empirical extensions of kinetic theory to low-density fluids,~\cite{1999_JPCM_11_5415, 1996_Nature_381_137, 2016_SciRep_6_20689, 2018_JCP_149_210901, 2019_PNAS_116_4070} and in the description of crystalline materials~\cite{Book_Born, Book_Venkataraman, 2021_CEC_23_5697} where a high degree of structural order simplifies the theoretical treatment. We thus might hope for a general relation between thermodynamic and transport properties in dense and cooled liquids as well, but this has not yet been demonstrated, even empirically.

The search for a relationship between the structural nature of matter and its dynamics has a long history. Ideas formulated long ago before the advent of modern science still permeate modern thinking and modeling of the dynamics of liquids. For example, the intuitive idea of ``free volume'' as a determinant of mobility, as formulated by Lucretius~\cite{Book_Lucretius, 1966_JCP_44_3950} over 2000 years ago, based on the analogy with the highly constrained motion operative in dense schools of fish, is the basis of a large and vibrant scientific literature on the dynamics of glass-forming (GF) fluids with passionate adherents.~\cite{1913_ZPC_84U_643, 1951_JAP_22_1031, 1950_JAP_21_581, 1951_JPC_55_221, 1959_JCP_31_1164, 1961_JCP_34_120, Book_Ferry, 1972_PNAS_69_3428} In a series of highly influential papers,~\cite{1963_JCP_39_3369, 1965_JCP_43_1852} Goldstein emphasized that this popular ``structural'' model of liquid dynamics could not adequately describe the pressure dependence of the dynamics of GF liquids, and he suggested that the attractive interactions between the molecules, which are necessary for the existence of the condensed state at constant pressure, should make either the enthalpy or the configurational entropy ($S_c$, the total entropy of a fluid minus the vibrational contribution) be the central thermodynamic property related to liquid dynamics.~\cite{1965_JCP_43_1852} Recent works have established the inadequacy of some more modern structurally-based models of liquid dynamics, such as the mode-coupling theory,~\cite{2010_PRE_82_031502} and other theories emphasizing the pair correlation function or averages of this quantity, providing molecular insights into the more abstract reasoning of Goldstein's energy landscape arguments. Parenthetically, we note that there are recent promising works to include information about higher-order density correlations into a mode-coupling framework, which appears to lead to a much improved agreement between predictions and simulations,~\cite{2015_PRL_115_205701, 2016_JSM_2016_054049, 2018_FP_6_97, 2022_PRL_Janssen} but historically fluid structure has often been identified with the readily observable pair correlation function or its Fourier transform, the static structure factor. Later, Goldstein~\cite{1965_JCP_43_1852} suggested that the configurational entropy might occupy a place of primacy in this type of relationship after the development of the Adam-Gibbs (AG) model~\cite{1965_JCP_43_139} of glass formation, and he first formulated his highly influential energy landscape view of the dynamics of GF liquids as an elaboration of this reasoning.~\cite{1969_JCP_51_3728} Notably, the configurational entropy contains information about density correlations through all orders~\cite{2012_JCP_137_024508, 2014_PRL_113_225701} so that models based on this property do not involve an approximation of structural correlations in terms of a pair correlation approximation.

The structure of the potential energy surface is governed by the same interplay of the relatively short-range hard-core repulsions of the molecules and the longer-range attractive interactions having various origins, depending on the molecular species that is reflected in the pair potential itself. Since molecules tend to become increasingly confined to the bottom of these potential wells as $T$ is lowered,~\cite{1996_PRE_54_964} the attractive contribution of the interparticle interactions generally makes a larger contribution to the dynamic properties of materials at lower temperatures, strongly modulating the dynamics of relaxation and diffusion in low-$T$ condensed liquids. This situation is contrasted with fluids at elevated temperatures~\cite{1996_PRE_54_964} where the dynamical motions are largely dictated by the unstable saddle point regions of the potential energy surface rather than regions near potential minima. Under these conditions, the moving particles in a low-density fluid can be reasonably modeled by ``effective'' hard spheres whose size depends somewhat on the form (e.g., power law) and strength (amplitude) of the pair potential between particles. Any successful model of the dynamics of liquids must bridge these high- and low-$T$ regimes in which the dynamics is insensitive and sensitive, respectively, to the landscape structure, while at the same time respecting the activated nature of relaxation and diffusion of materials in the equilibrium condensed state at any finite $T$.

We should mention that Sutherland~\cite{1890_PM_30_318, 1891_PM_32_31, 1891_PM_32_215, 1891_PM_32_524} formulated a pioneering kinetic theory of solids not long after Maxwell's kinetic theory of gases,~\cite{1867_PTRS_157_49} which emphasized that the rigidity of matter should be understood, as in the case of gases, as a kinetic rather than structural interpretation emphasized by the ``great French elasticians''. Based on this theory, Sutherland deduced that ``liquefaction occurs when each molecule ceases to be hemmed in and just manages to wriggle through amongst its neighbors'', and he quantified this physical picture by arguing that the melting condition corresponds to the ratio of the volume explored by the particle relative to the molecular size achieved, this ratio being independent of the substance. This is apparently the first formulation of the semi-empirical Lindemann criterion in a particularly sophisticated form. A discussion of the historical development of the Lindemann relation can be found in ref~\citenum{2021_Mac_54_3247}. This rattle volume, which is a kind of dynamical free volume, is central to recent modeling of the relaxation of GF liquids, as we shall discuss below. In this series of papers, Sutherland~\cite{1905_PM_9_781} was also the first to introduce what is now widely referred to as the Stokes-Einstein relation describing the dependence of the diffusion coefficient of molecules on temperature and viscosity under homogeneous fluid conditions, where his predictions also included a consideration of hydrodynamic slip at the boundary of the molecules. In one of the first scientific contributions to polymer physics, Onsager~\cite{1932_PR_40_1028} also emphasized the importance of partial slip in the hydrodynamics of polymers in solutions. While partial slip is normally ignored in polymer hydrodynamic modeling, the rationale for assuming a stick boundary condition on molecules remains unclear. These works are among the earliest works to apply macroscopic hydrodynamics to molecules, an approach that has been highly fruitful scientifically as a model of molecular dynamics, but which is not without its difficulties. Our main point here is that Sutherland was far ahead of his time, which probably explains his current lack of appreciation today.

The energy landscape conception of liquid dynamics was greatly elaborated by Stillinger and Weber,~\cite{1982_PRA_25_978, 1984_Science_225_983} and this approach has led to accurate numerical methods for precisely calculating the configurational entropy of fluids and in conceptualizing the dynamics of condensed materials.~\cite{2005_JSM_2005_P05015, 2001_Nature_409_164, 1998_Nature_393_554, 2008_JPCM_20_373101, 2018_ARPC_69_401} Goldstein's work,~\cite{1963_JCP_39_3369} along with the earlier lattice model of polymer glass formation emphasizing the configurational entropy $S_c$ by Gibbs and DiMarzio,~\cite{1958_JCP_28_373} provided an impetus for the pioneering work of Bestul and Chang,~\cite{1964_JCP_40_3731} who first established a strong empirical correlation between the excess entropy $S_{\mathrm{exc}}$ (the entropy of a material in the fluid state minus that of the crystal or glass form of the material in its ``solid'' state) and the rate of structural relaxation in GF liquids. Shortly thereafter, the AG model~\cite{1965_JCP_43_139} of relaxation in GF liquids was formulated in terms of a hybridization of transition state theory (TST)~\cite{1941_CR_28_301, Book_Eyring} and an emphasis on $S_c$. Recent works by Caruthers and Medvedev~\cite{2018_PRM_2_055604, 2019_Mac_52_1424} have brought renewed attention to the original suggestion of Goldstein~\cite{1963_JCP_39_3369} that the fluid enthalpy should have a central significance in understanding the dynamics of cooled liquids, and the merit of this proposal has been carefully reviewed recently by Zhao and Simon.~\cite{2020_JCP_152_044901} We also point out important works of Johari,~\cite{2000_JCP_112_8958, 2000_JCP_112_7518} which discuss the many and convoluted difficulties of estimating $S_c$ experimentally, a problem that extends to estimation of the configurational enthalpy and internal energy of fluids. Basically, it is difficult, if not impossible, to precisely estimate experimentally the vibrational contribution~\cite{2000_JCP_112_8958, 1976_JCP_64_4767} to the thermodynamic properties of cooled polymeric liquids to enable the precise estimation of the fluid configurational properties as the difference between the corresponding property and the vibrational contribution. Estimating this difference is also made difficult because the vibrational contribution can constitute the largest contribution to this thermodynamic property. This fact has been demonstrated from explicit computational estimates of $S_c$ of a coarse-grained polymer melt by molecular dynamics (MD) simulations that allow for the precise estimation of $S_c$ and the total fluid entropy.~\cite{2013_JCP_138_12A541} It should then be no surprise that the extrapolation of $S_{\mathrm{exc}}$ to zero at low $T$ often occurs in polymeric GF liquids at a rather different temperature than the temperature $T_0$ deduced from the corresponding divergence of the relaxation time and viscosity fitted to the Vogel-Fulcher-Tammann (VFT) relation.~\cite{1921_PZ_22_645, 1925_JACS_8_339, 1926_ZAAC_156_245} For example, Miller reported a discrepancy between these temperatures as large as $60$ K in polystyrene,~\cite{1970_Mac_3_674} and similar discrepancies have been noted for many other polymer materials.~\cite{2005_EPL_70_614} We expect that the unsatisfactory nature of the approximation $S_c \approx S_{\mathrm{exc}}$ for polymer fluids is responsible for the utter failure of thermodynamic fragility estimates of Martinez and Angell~\cite{2001_Nature_410_663} to predict even the correct \textit{qualitative} trend when these estimates are applied to polymer materials.~\cite{2001_JCP_114_5621}

We also mention interesting recent works of Schweizer and coworkers,~\cite{2021_Mac_54_10086, 2021_PNAS_118_e2025341118} which emphasizes the relation between the isothermal compressibility $\kappa_T$ (the reciprocal of the bulk modulus, $B$) as being the thermodynamic property of prime significance in relation to understanding the rate of relaxation in GF liquids, an idea suggested in earlier modeling of diffusion in crystalline materials~\cite{1981_PRB_24_904, 1980_PRB_22_3130, 2009_PRB_79_132204} and in a variant of the mode-coupling theory of the dynamics of GF liquids.~\cite{1990_ZPB_79_5}

Collectively, the growing number of proposed models attempting to quantitatively ``relate'' the thermodynamics to the dynamics of liquids would seem to indicate an increasing lack of consensus on the nature of glass formation, but we see these proliferating models in a different way. The fact that so many models of glass formation lead to such a common phenomenology suggests to us instead that there might be some hidden unity underlying these models, an optimistic hypothesis that underlies the present work. To demonstrate this explicitly, we utilize the lattice cluster theory (LCT)~\cite{1998_ACP_103_335, 2014_JCP_141_044909, 2021_Mac_54_3001} to examine the extent to which the configurational entropy, enthalpy, and internal energy are interrelated in a wide range of model GF polymer liquids having variable architecture, chain length, cohesive interaction strength, and chain rigidity and under different applied pressures. Importantly, such a study is not limited to any particular model of how these thermodynamic properties might be related to the dynamics of these liquids. We qualitatively infer from the recent measurements of Caruthers and Medvedev~\cite{2018_PRM_2_055604, 2019_Mac_52_1424} and Zhao and Simon~\cite{2020_JCP_152_044901} that these thermodynamic properties must be strongly interrelated, thus potentially offering an opportunity for some unification of the models of glass formation. Because the LCT by construction excludes consideration of the vibrational contribution to the free energy of polymer fluids, this theory allows for the calculation of configurational properties of polymeric fluids without approximation than those inherent in the mean-field thermodynamic theory, so that our analysis is not plagued by uncertainties in estimates of configurational thermodynamic properties. Our analysis indeed reveals the close interrelation between these thermodynamic properties so that links are established between the respective models of glass formation. The present work can be viewed as an extension of a previous work attempting to obtain some unification of the various models of glass formation based on MD simulations.~\cite{2015_PNAS_112_2966} We utilize this approach to help augment our discussion based on the LCT and observations drawn from the generalized entropy theory (GET),~\cite{2021_Mac_54_3001, 2008_ACP_137_125} a combination of the LCT~\cite{1998_ACP_103_335, 2014_JCP_141_044909, 2021_Mac_54_3001} and the AG model.~\cite{1965_JCP_43_139}

We mention another thermodynamic aspect of the dynamics of GF liquids, which serves a useful role in discriminating amongst the various proposed models. Both the thermodynamic and dynamic properties of fluids with particles interacting through power-law potentials have been observed to obey a non-trivial scaling as a function of the volume $V$ raised to a power $\gamma$ times $T$, $T V^{\gamma}$, a phenomenon termed ``thermodynamic scaling''.~\cite{2005_RPP_68_1405, 2010_Mac_43_7875, Book_Roland, Book_Paluch, 1971_JCP_55_1128} This scaling can be rigorously derived for model liquids at constant density composed of particles interacting with a spherically symmetric power-law or ``soft sphere'' intermolecular potential,~\cite{2009_JCP_131_234504, 2019_JCP_151_204502} $U(r) \sim |r|^{-\alpha}$, where $|r|$ is the interparticle distance and $\gamma = d / \alpha$ with $d$ being the spatial dimension. This \textit{analytic symmetry} appears to generally describe experimentally the $T$ and $\rho$ dependence of relaxation and diffusion in liquids, although it is not clear that this general scaling derives from the intermolecular potential being a homogeneous function, as in the model just mentioned.~\cite{2021_Mac_54_3247} In our discussion below, we will consider whether the thermodynamic properties considered are consistent with this type of scaling as a ``filtering'' criterion~\cite{2009_JCP_131_234504} for deciding whether a given thermodynamic property is a viable candidate for being related to the fluid dynamics based on this empirical criterion. As a general matter, we interpret the existence of thermodynamic scaling as qualitatively supporting the hypothesis that the thermodynamics and dynamics of liquids are related, but this relation is not diagnostic as to any particular interrelation.

The organization of our paper involves three thrusts that are closely interrelated. We first utilize the LCT for the thermodynamic properties of polymer fluids to show that the configurational entropy, enthalpy, and internal energy are indeed closely interrelated, as suggested by recent measurements by Caruthers and Medvedev.~\cite{2018_PRM_2_055604, 2019_Mac_52_1424} This finding is independent of any question of how these thermodynamic properties might be related to the dynamics of liquids and thus is more easily addressed. Based on these results, we show that the predictions of the GET of glass formation can be recast in terms of any of these thermodynamic properties as a matter of \textit{convenience}. Alternatively, we can cast the results of the GET in terms of temperature and molecular parameters, which we have generally preferred in the past because of inherent ambiguities in estimating configurational thermodynamic properties experimentally.

As a second stage of our primarily thermodynamically oriented analysis, we then consider the occurrence of ``thermodynamic scaling'' to assess which thermodynamic properties are most naturally linked to the dynamics of liquids in the specific sense of exhibiting this scaling symmetry. We utilize both the GET and MD simulations of a coarse-grained model of polymer melts having variable rigidity and pressure to explore the origin of this apparently general scaling property of the dynamics of fluids. Although thermodynamic scaling is found to be a powerful tool for discriminating which thermodynamic properties are more correlated with changes in the dynamics, we then encounter the interesting question of why some basic properties that one might most expect to exhibit this scaling, such as the isothermal compressibility, do not exhibit this scaling property. As a general matter, we then interpret the existence of thermodynamic scaling as qualitatively supporting the hypothesis that the thermodynamics and dynamics of liquids are closely interrelated, but this currently empirical relationship is still somewhat mysterious in its origin. We are also struck by the fact that this scaling property is clearly exhibited by the reduced configurational entropy density,~\cite{2021_Mac_54_3247} defined by the configurational entropy normalized by its value at the onset temperature $T_A$ below which the dynamics start to deviate from the simple Arrhenius behavior, which means that the GET is consistent with this scaling symmetry. Other models of the dynamics of fluids based on thermodynamic properties that can be related to this basic thermodynamic property also cannot be excluded from consideration. We thus have a criterion for ``filtering'' models of glass formation, even if we do not fully comprehend the physical origin of thermodynamic scaling.

The second thrust of our paper is based on the recognition that the configurational entropy involves just a consideration of the total number of accessible energy minima in the free energy surface as a function of temperature or other thermodynamic variables, such as $P$, and thus quantifies the overall ``complexity'' of this energy surface in this sense. The variation of the configurational entropy with temperature or molecular parameters provides information about the average ``roughness'' of the topography of these energy surfaces as viewed relative to some average energy level that the system explores at thermal equilibrium. An understanding of the dynamics of liquids at equilibrium requires an understanding of transitions between accessible minima in the energy surface driven by thermal fluctuations, which requires other metrical information about the energy landscape that is not obviously encompassed by any thermodynamic theory. In concrete terms, the GET requires information about the activation free energy parameters (i.e., enthalpy $\Delta H_o$ and entropy $\Delta S_o$ of activation) in the high temperature Arrhenius regime that formally exist in any general transition state theory (TST) framework.~\cite{1941_CR_28_301, Book_Eyring} The lack of a direct method for calculating $\Delta H_o$ and $\Delta S_o$ is the weakest link of the GET model in relation to the prediction of relaxation in real materials.~\cite{2021_Mac_54_3001} We have approached this difficult problem, as many others did in the past in modeling condensed materials,~\cite{1995_PRL_75_469, 2001_PRB_64_075418, 2011_PNAS_108_5174} through the determination of $\Delta H_o$ and $\Delta S_o$ from MD simulations or experimental measurements based on TST. In the present paper, we describe recent computational efforts to quantify the complex saddle point geometry of energy surfaces to understand how the activation free energy parameters might also be interpreted in terms of the geometry of the energy landscape and how the metrical structure of the energy landscape is related to the thermodynamics of the fluid. This is a necessary link if the dynamics of the fluid is to be related to its thermodynamic properties. Ultimately, we would also like to understand how the geometry of these energy surfaces engenders the propensity for collective motion in cooled liquids that would allow for a direct understanding or modification of the premises on which the AG model~\cite{1965_JCP_43_139} is based. This work is ongoing, but there has been progress on this fundamental problem that can guide future work in testing tentative ideas about the general metrical structure of energy landscapes and how this structure gives rise to links with thermodynamic properties.

We also recognize that the $\alpha$-relaxation process or segmental relaxation time $\tau_{\alpha}$ in polymer fluids is just a narrow aspect of glass formation so that we must integrate progress in understanding this relaxation process with other basic relaxation process such as the slow Johari-Goldstein (JG) $\beta$-relaxation process,~\cite{1970_JCP_53_2372, 1971_JCP_55_4245, 1973_JCP_58_1766} which has prominent significance in materials in their glass state where the $\alpha$-relaxation time is too long to appreciably contribute to relaxation and diffusion. In our third thrust, we thus show that MD simulations allow us to leverage our understanding of $\tau_{\alpha}$ to establish definite and apparently general scaling relations to the $\beta$-relaxation process. We view this type of effort as another essential element in developing an integrated understanding of all aspects of the phenomenon of glass formation. Our analysis so far appears to be broadly consistent with the hypothesis of the existence of deep interrelations between the dynamics and thermodynamics of GF liquids, a thread that we continue to follow in new directions.

\section{Methods}

\subsection{\label{Sec_GET}Generalized Entropy Theory}

We have recently reviewed the entropy theory approach to modeling the dynamics of GF liquids, and the GET model in particular,~\cite{2021_Mac_54_3001} where the basic ideas underlying the model and the key predictions are extensively discussed. The GET combines the LCT,~\cite{1998_ACP_103_335, 2014_JCP_141_044909} a model of polymer thermodynamics that extends the Flory-Huggins theory~\cite{1941_JCP_9_660, 1941_JCP_9_440, 1942_JPC_46_151} to include a description of variable monomer structure, interaction, and chain rigidity, and the AG relation~\cite{1965_JCP_43_139} to allow for the investigations of the influence of basic molecular parameters on the thermodynamics and segmental dynamics of polymers undergoing glass formation. The LCT, which is the purely thermodynamic component of this approach, describes polymers in terms of a set of united-atom groups that are placed on a hypercubic lattice with a total number of $N_l$ lattice sites, and each united-atom group occupies a single lattice site of volume $V_{\mathrm{cell}}$ so that the volume of the system is $V = N_l V_{\mathrm{cell}}$. The lattice model also includes empty sites that are not occupied by united-atom groups, and hence, the theory allows for the computations of properties under different applied $P$. The discretization of the fluid structure by a lattice description is an advantage in the sense that this procedure allows for the systematic computation of almost any thermodynamic property of interest, but this approximation is also a weakness because it limits the physical faithfulness in the description of the true atomic structure.

The LCT allows us to derive an analytic expression for the Helmholtz free energy $F$ as a function of temperature $T$, polymer filling fraction $\phi$, molecular mass $M$, microscopic cohesive energy parameter $\epsilon$, bending energy parameter $E_b$, and a set of geometrical indices that reflect the size, shape, and bonding patterns of the monomers. For the explicit expression of the free energy and the meaning of the parameters, see refs~\citenum{1998_ACP_103_335} and~\citenum{2014_JCP_141_044909} for a discussion. The common thermodynamic quantities are then readily obtained; e.g., the entropy, internal energy, and enthalpy are calculated from their standard definitions,
\begin{equation}
	\label{Eq_Thermo}
	S_c = - \left. \frac{\partial F}{\partial T} \right|_{\phi}, \ U_c = \left. \frac{\partial [F / (k_{\mathrm{B}}T)]}{\partial [1 / (k_{\mathrm{B}}T)]} \right|_{\phi}, \ H_c = U_c + PV,
\end{equation}
where $k_{\mathrm{B}}$ is Boltzmann's constant. Here, we have used the subscript ``$c$'' to denote that the thermodynamic properties are configurational in origin. In the following, we focus on these properties normalized by the number of lattice sites, namely, the configurational entropy density $s_c$, configurational internal energy density $u_c$, and configurational enthalpy density $h_c$,
\begin{equation}
	\label{Eq_ThermoNorm}
	s_c = S_c / N_l,\ u_c = U_c / N_l,\ h_c = H_c / N_l.
\end{equation}
The cohesive energy density is related to the configurational internal energy density as
\begin{equation}
	\label{Eq_PICED}
	\Pi_{\mathrm{CED}} = |U_c| / V = |u_c| / V_{\mathrm{cell}}.
\end{equation}
In the present work, the cell volume parameter is selected to be $V_{\text{cell}} = 2.5^3 \text{\AA}^3$.

The GET provides a predictive framework for computing the characteristic properties of polymer glass formation. The ``onset temperature'' $T_A$ signals the onset of non-Arrhenius behavior of $\tau_{\alpha}$ and is determined by a temperature corresponding to the maximum $s_c^*$ of $s_c(T)$. The crossover temperature $T_c$ separates two regimes of $T$ with qualitatively different dependences of $\tau_{\alpha}$ on $T$, as discussed below, and this temperature is estimated from $\partial^2 (Ts_c)/\partial T^2 = 0$. The determination of the glass transition temperature $T_{\mathrm{g}}$ follows its operational definition based on the condition, $\tau_{\alpha}(T_{\mathrm{g}}) = 100$ s. The GET computes $\tau_{\alpha}$ from the AG relation,~\cite{1965_JCP_43_139}
\begin{equation}
	\label{Eq_AG}
	\tau_{\alpha} = \tau_o \exp\left[ z(T) \Delta G_o / k_{\mathrm{B}}T \right],\ z(T) = s_c^*/s_c(T),
\end{equation}
where $z(T)$ corresponds to the extent of collective motion, or more specifically, the number of segments in the abstract ``cooperatively rearranging regions'' (CRR) of the AG model.~\cite{1965_JCP_43_139} The GET assumes that the high temperature vibrational prefactor of polymer materials takes a value of $\tau_o = 10^{-13}$ s, a typical experimental estimate for polymers.~\cite{2003_PRE_67_031507} $\Delta G_o$ is the activation free energy at high $T$, which is anticipated from TST~\cite{1941_CR_28_301, Book_Eyring} to contain both enthalpic $\Delta H_o$ and entropic $\Delta S_o$ contributions, i.e., $\Delta G_o = \Delta H_o - T \Delta S_o$. The physical origin of the Arrhenius temperature dependence observed at high temperatures was first seriously considered by Raman~\cite{1923_Nature_111_532} and Andrade,~\cite{1930_Nature_125_309} whose works are still well worth reading because of their insights into the nature of the liquid state. Ewell~\cite{1938_JAP_9_252} discussed numerous earlier theoretical works associated with the development of an understanding of Arrhenius diffusion and viscosity variations with temperature and with phenomenology that we now recognize as being described by TST.

Motivated by the heuristic approximation made by AG~\cite{1965_JCP_43_139} that the entropic contribution to the activation free energy is negligible, i.e., $\Delta S_o = 0$, along with the simulation evidence for a limited number of coarse-grained liquid models and some experimental evidence for model liquids indicating the approximation,~\cite{2008_ACP_137_125} $\Delta H_o \approx (7 \pm 1) k_{\mathrm{B}} T_c$, the original GET model simply assumed that $\Delta G_o = 6k_{\mathrm{B}} T_c$ for a rough estimation of the structural relaxation time of polymers having complex structure with no other parameters other than those required to describe the thermodynamics of the polymer material. As described above, the crossover temperature $T_c$ in the GET separates the high- and low-$T$ regimes of glass formation and is precisely defined by an inflection point in the product of $s_c$ times $T$.~\cite{2008_ACP_137_125} The structural relaxation time is predicted by the GET to exhibit an apparent power-law temperature dependence near $T_c$, $\tau_{\alpha} \sim (T - T_c)^{-\gamma_c}$ with $\gamma_c$ being a crossover exponent, providing an experimental criterion for locating $T_c$ experimentally. In Section~\ref{Sec_Implication}, we discuss a possible origin of this relation between the activation energy $\Delta H_o$ and $T_c$, the key relation linking the dynamics and thermodynamics of dense fluids within the GET framework, from an energy landscape perspective. 

Experimentalists often correlate the ``activation energy'' $\Delta H_o$ with $T_{\mathrm{g}}$.~\cite{2012_PRE_86_041507, 2015_Mac_48_3005} Since the characteristic temperature ratio $T_c / T_{\mathrm{g}}$ in polymer fluids is normally in the range $1.2$ to $1.3$,~\cite{2008_ACP_137_125} our computational criterion accords with the experimental correlation between $\Delta H_o$ and $k_{\mathrm{B}} T_{\mathrm{g}}$ for molecular fluids within the broadly stated uncertainties just mentioned, $\Delta H_o \approx (10 \pm 1) k_{\mathrm{B}} T_{\mathrm{g}}$ (see Figure 4 in ref~\citenum{2012_PRE_86_041507}). Somewhat higher values of the prefactor in this relation have been reported for large molecular fluids, $\Delta H_o \approx 16 k_{\mathrm{B}} T_{\mathrm{g}}$. The estimation of $\Delta H_o$ is notoriously difficult in polymer materials because of limited thermal stability at very high $T$ so that there is a tendency to overestimate $\Delta H_o$.~\cite{1997_Polymer_38_1477} Despite these issues, we think that the estimates of the mass dependence of $\Delta H_o$ by R{\"o}ssler and coworkers~\cite{2015_Mac_48_3005} are convincing so that these results require serious consideration. In small-molecule liquids, $\Delta H_o$ is often estimated rather effectively from the heat of vaporization of the fluid.~\cite{Book_Eyring, 1938_JAP_9_252} This common, and widely utilized, semi-empirical expression in Eyring's TST~\cite{1941_CR_28_301, Book_Eyring} has its origin in long-standing parallels in the $T$ dependence of the dynamics and the vapor pressure of liquids that go back nearly a century.~\cite{1934_JAP_5_39} This is just another example of the recognition of the parallelism between the dynamics and thermodynamics of liquids that continues to inspire correlative relationships. Egami and coworkers~\cite{2013_PRL_110_205504} have revived this picture of elementary particle displacements occurring as a kind of local ``evaporation'' process, as envisioned even earlier by Frenkel.~\cite{Book_Frenkel} In particular, they physically interpreted the structural relaxation time of the fluid in terms of local thermal ``excitation'' events in which a particle in the local coordination sphere displaces to a distance on the order of a particle diameter. In the specific model of this type of excitation process governing molecular diffusion and relaxation, Eyring and coworkers~\cite{1941_CR_28_301, Book_Eyring} modeled the particle displacement to occur as either a unimolecular or bimolecular reaction process in which the potential energy change involved in the postulated thermal activated process determines the activation energy. In particular, the activation energy as inferred to equal the heat of vaporization of the liquid (an accessible measure of the cohesive interaction strength for non-polymeric liquids) divided by a factor of $3$ or $4$ was argued to be related to reaction order. This evaporation picture of activated transport in liquids serves to rationalize the close empirical relation between $\Delta H_o$ and the cohesive energy density observed in the dynamics of many small-molecule liquids,~\cite{1997_FPE_140_221, 2003_AIChEJ_49_799} including ionic liquids.~\cite{2009_JPCB_113_12353} Kuazmann and Eyring~\cite{1940_JACS_62_3113} discussed the limitations of this model of the activation energy in high molecular polymers, where it was reasonably argued that the displacing molecular segments are molecular segments rather than the polymer as a whole when the polymer mass becomes large. Accordingly, this qualitative physical picture of the activation energy in polymer melts was adapted into quantitative models~\cite{1988_PMA_57_217, 1994_TA_238_41} in which the cohesive energy density of the fluid plays a central role.

Of course, there are many properties that increase with the cohesive energy density, such as the melting temperature, enthalpy of fusion, boiling temperature, and critical temperature for liquid-vapor transformation. Then almost any thermodynamic property related to cohesive energy density should serve the purpose of establishing a correlative relation with $\Delta H_o$, at least for a limited class of substances. Various other proposed properties of this kind have been discussed by Zhang et al.~\cite{2021_JCP_155_174901} in an effort to understand mobility gradients in supported polymer films. For example, a correlation between $T_{\mathrm{g}}$ and the melting temperature $T_m$ is also very popular for materials that readily crystallize.~\cite{1950_JAP_21_1189, 1980_AM_28_1085} Caruthers and Medvedev~\cite{2018_PRM_2_055604} suggested that the activation parameter $B$ in their model could be estimated from the heat of fusion. More generally, it should be possible to determine $\Delta H_o$ from high-$T$ atomistic simulations or from experiments in favorable circumstances, and this property would make a good topic for a machine-learning-based study, given that this quantity can be obtained from relatively high-$T$ simulations.

The reasoning for the relationship between $\Delta H_o$ and $T_c$ noted above has been discussed at length in ref~\citenum{2008_ACP_137_125}, and we refer the reader to this reference for details. Importantly for the present discussion, simulation results have indicated that the entropic contribution to $\Delta G_o$ is generally not negligible in GF polymers,~\cite{2014_NatCommun_5_4163, 2015_PNAS_112_2966, 2015_JCP_142_234907, 2016_Mac_49_8355, 2016_MacroLett_5_1375, 2017_Mac_50_2585, 2020_Mac_53_4796, 2020_Mac_53_6828, 2020_Mac_53_9678, 2021_Mac_54_9587, 2022_Mac_55_3221} where a strong correlation between $ \Delta H_o$ and $\Delta S_o$ has been found. We discuss below why such correlations should generally be expected in both the thermodynamics and dynamics of condensed materials. Nevertheless, we believe that the simplifying assumption of the AG and GET models, $\Delta S_o = 0$, should be sufficient to capture the general trends of polymer glass formation when molecular or thermodynamic parameters are varied, as indicated by previous simulations for a range of polymer systems.~\cite{2020_Mac_53_4796, 2020_Mac_53_6828, 2020_Mac_53_9678, 2021_Mac_54_3247, 2021_Mac_54_6327, 2022_Mac_55_3221} However, we have also found that the inclusion of $\Delta S_o$ is important for the \textit{quantitative} description of the relaxation times of particular materials based on the GET, which is our ultimate goal. In our recent review of the GET and entropy theory of glass formation,~\cite{2021_Mac_54_3001} we have discussed how the inclusion of $\Delta S_o$ into a revised GET framework improves the fit between the model and experiment. 

\begin{figure*}[htb]
	\centering
	\includegraphics[angle=0,width=0.4\textwidth]{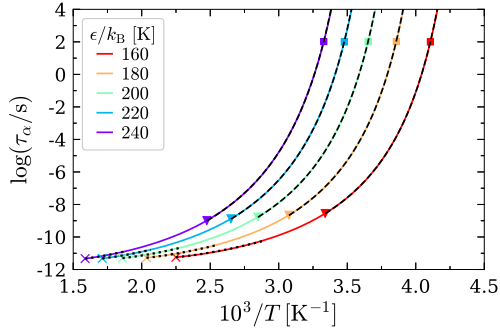}
	\caption{\label{Fig_GET_Tau} Logarithm of the structural relaxation time, $\log \tau_{\alpha}$, as a function of inverse temperature $10^3/T$ calculated from the generalized entropy theory (GET) for varying cohesive energy parameters $\epsilon$ at zero pressure for polymer melts with the structure of polypropylene (PP) having a chain length of $N_c = 8000$ and a bending energy parameter of $E_b/k_{\mathrm{B}} = 600$ K. The cross, triangle, and square symbols indicate the positions of the onset $T_A$, crossover $T_c$, and glass transition temperatures $T_{\mathrm{g}}$ of glass formation, respectively. Dashed and dotted lines correspond to the descriptions based on eqs~\ref{Eq_VFT} and~\ref{Eq_HighT}, respectively.}
\end{figure*}

The characteristic $T_0$, sometimes termed the ``ideal glass transition temperature'', and the fragility index $D$ can be obtained from the Vogel-Fulcher-Tammann (VFT) equation,~\cite{1921_PZ_22_645, 1925_JACS_8_339, 1926_ZAAC_156_245}
\begin{equation}
	\label{Eq_VFT}
	\tau_{\alpha} = \tau_0\exp\left(\frac{DT_0}{T-T_0}\right),
\end{equation}
where $\tau_0$ is a prefactor on the order of vibrational relaxation time, i.e., $10 ^{-14}$ s--$10^{-13}$ s.~\cite{1997_Polymer_38_6261} It is emphasized that the VFT equation directly follows from the GET with the parameters $D$ and $T_0$ exactly determined by direct computations in terms of molecular parameters, where this expression is predicted to \textit{only apply} over a $T$ range above $T_{\mathrm{g}}$, but below the ``crossover temperature'' $T_c$ separating the high and low temperature regimes of glass formation.~\cite{2008_ACP_137_125} The VFT equation has been noted to apply in previous experimental studies over a similar $T$ range.~\cite{1955_JACS_77_3701} The use of the VFT expression does not imply, however, that $\tau_{\alpha} $ actually diverges at $T_0$, as recently discussed at length in ref~\citenum{2021_Mac_54_3001}. The GET is agnostic on the question of whether $\tau_{\alpha}$ actually diverges at $T_0$ in a temperature range beyond the scope of this model.~\cite{2021_Mac_54_3001}

Figure~\ref{Fig_GET_Tau} shows $\log \tau_{\alpha}$ versus inverse temperature for varying cohesive energy parameters $\epsilon$, along with the VFT fits. The calculations are performed for polymer melts with the structure of polypropylene (PP) at zero pressure, where the chain length is $N_c = 8000$ and the bending energy parameter is $E_b/k_{\mathrm{B}} = 600$ K. The GET also allows us to compute the ``fragility index'' or ``steepness index'' proposed by Angell,~\cite{1991_JNCS_131_13}
\begin{equation}
	\label{Eq_Angell}
	m \equiv \left. \frac{\partial \log\tau_{\alpha}}{\partial (T_{\mathrm{g}}/T)} \right|_{T=T_{\mathrm{g}}}.
\end{equation}
Since the VFT equation holds to a good approximation near $T_{\mathrm{g}}$, $m$ is related to $D$ via $m = DT_{\mathrm{g}}T_0 / [(T_{\mathrm{g}}-T_0)^2\ln 10]$. 

While the VFT relation describes well the $T$ dependence of $\tau_{\alpha}$ predicted from the GET in the low-$T$ regime of glass formation, a different functional form provides a better description of the $T$ dependence of $s_c$ and thus $\tau_{\alpha}$ in the high-$T$ regime of glass formation,~\cite{2021_Mac_54_3001, 2008_ACP_137_125}
\begin{equation}
	\label{Eq_HighT}
	s_c^*/s_c - 1 = C_s [(T_A - T)/T_A]^2,\ T_c < T_A - 100\ \mathrm{K} < T < T_A,
\end{equation}
where $C_s$ measures the steepness of the $T$ dependence of $s_c$. The description based on eq~\ref{Eq_HighT} is shown as dotted lines in Figure~\ref{Fig_GET_Tau}. $C_s$ can be viewed as a measure of fragility in the high-$T$ regime of glass formation where the VFT equation is \textit{not} valid according to the GET. We note that eq~\ref{Eq_HighT} in junction with the GET expression for $\tau_{\alpha}$ (eq~\ref{Eq_AG}) is exactly equivalent to the ``modified parabolic'' model of the structural relaxation discussed by Egami and coworkers,~\cite{2015_SciRep_5_13837} where the parameters of this model expression for $\tau_{\alpha}$ can also be calculated directly from the GET rather than obtained by curve-fitting to measurements. The $T$ range in which this $T$ expression of $\tau_{\alpha}$ is predicted to apply is between the onset temperature $T_A$ for non-Arrhenius relaxation and the crossover temperature $T_c$, the high-$T$ regime of glass formation. The quantity $C_s$ should be of great interest from a simulation viewpoint since simulations are mostly restricted to temperatures higher than $T_c$ in GF liquids because of the very long relaxation times at lower $T$. Equation~\ref{Eq_HighT} has been shown to hold in simulations of a coarse-grained polymer melt with and without antiplasticizer additives.~\cite{2007_JCP_126_234903} As another merit, eq~\ref{Eq_HighT} provides a useful method for estimating $T_A$ in simulations.~\cite{2009_PNAS_106_7735, 2016_JSM_054048, 2020_JCP_153_124508} We plan to investigate eq~\ref{Eq_HighT} and the alternative ``fragility'' parameter $C_s$ of the high-$T$ regime of glass formation more thoroughly in future simulation studies. It would also be of great interest to explore the relation between $C_s$ and the fragility parameters of the low-$T$ regime of glass formation, such as $m$ and $D$ discussed above.

We note that all the characteristic temperatures of glass formation ($T_A$, $T_c$, $T_{\mathrm{g}}$, and $T_0$) can be directly calculated from the GET,~\cite{2021_Mac_54_3001, 2008_ACP_137_125} along with the corresponding characteristic relaxation times at these temperatures precisely determined. The GET model is highly predictive once the variables governing the thermodynamics of the material have been determined.

Here, we make additional comments on the choices of $\tau_o$ and $\Delta H_o$ in the GET. If we take $\tau_o$ to be on the order $10^{-14}$ and $\Delta H_o$ to be the somewhat revised value of $7 k_{\mathrm{B}} T_c$, which are more consistent with the range of values of these parameters suggested previously, then $\tau_{\alpha}$ at $T_A$ and $T_c$ are predicted to be about $10^{-12}$ s and $10^{-10}$ s--$10^{-6}$ s for the molecular models considered in the present paper. Then $\tau_{\alpha}$ at $T_A$ is on the order of the fast $\beta$-relaxation time,~\cite{2018_JCP_148_104508, 2003_PRE_67_031507} a natural result that arises because the $\alpha$-relaxation time bifurcates from the $\beta$-relaxation process near $T_A$. A previous work has recognized that $\tau_{\alpha}$ at $T_c$ takes a ``magic value'' in the limited range, $10^{-7 \pm 1}$ s.~\cite{2003_PRE_67_031507} These characteristic relaxation times are then ``quasi-universal'' in that they have a typical order of magnitude, as in the case of $\tau_{\alpha}$ at $T_{\mathrm{g}}$, which is often defined by the condition $\tau_{\alpha} = 100$ s. The high-$T$ regime of glass formation between $T_A$ and $T_c$ thus corresponds to a change of $\tau_{\alpha}$ by about $4$ orders of magnitude, and in the low-$T$ regime, where the VFT equation is predicted from the GET to apply, $\tau_{\alpha}$ changes by about $10$ orders of magnitude. In addition to specific predictions for particular materials, we also see that the GET gives insight into general trends in glass formation of molecular fluids that are largely independent of the chemical nature of the material.

\subsection{\label{Sec_MD}Molecular Dynamics Simulation}

Our simulation study is based on a coarse-grained bead-spring model of polymers,~\cite{1990_JCP_92_5057, 1986_PRA_33_3628} where the chains are represented by a certain number of connected statistical segments (beads). Notably, the bead is generally not equivalent to the Kuhn segment, and the Kuhn length $l_K$ must be measured in simulations.~\cite{2020_Mac_53_1901} Neighboring beads along a chain are connected by the finitely extensible nonlinear elastic (FENE) potential,~\cite{1990_JCP_92_5057, 1986_PRA_33_3628}
\begin{eqnarray}
	\label{Eq_FENE}
	U_{\mathrm{FENE}}(r) = -\frac{1}{2} k_b R_0^2 \ln\left[1 - \left(\frac{r}{R_0}\right)^2\right] + 4 \varepsilon \left[ \left(\frac{\displaystyle \sigma}{\displaystyle r}\right)^{12} - \left(\frac{\displaystyle \sigma}{\displaystyle r}\right)^6 \right] + \varepsilon,
\end{eqnarray}
where $r$ denotes the distance between two beads and $\varepsilon$ and $\sigma$ are the energy and length scales associated with the Lennard-Jones (LJ) potential. The first term of eq~\ref{Eq_FENE} extends to $R_0$, and the second term has a cutoff at $2^{1/6} \sigma$. We use common choices for the parameters $k_b$ and $R_0$, namely, $k_b = 30 \varepsilon/\sigma^2$ and $R_0 = 1.5 \sigma$. Interactions between all nonbonded pairs are described by a truncated-and-shifted LJ potential,
\begin{eqnarray}
	\label{Eq_LJ}
	U_{\mathrm{LJ}}(r) = 4 \varepsilon \left[ \left(\frac{\displaystyle \sigma}{\displaystyle r}\right)^{12} - \left(\frac{\displaystyle \sigma}{\displaystyle r}\right)^6 \right] + C(r_{\mathrm{cut}}),\ r < r_{\mathrm{cut}},
\end{eqnarray}
where $C(r_{\mathrm{cut}})$ is a constant to ensure that $U_{\mathrm{LJ}}$ varies smoothly to zero at the cutoff distance $r_{\mathrm{cut}}$. We choose $r_{\mathrm{cut}} = 2.5\sigma$ to include attractive nonbonded interactions. Chain rigidity is controlled by an angular potential,~\cite{2019_JCP_150_091101}
\begin{eqnarray}
	\label{Eq_Bend}
	U_{\mathrm{bend}}(\theta) = -A \sin^2(B \theta),\ 0 < \theta < \pi / B,
\end{eqnarray}
where the bond angle is given by $\theta = \cos^{-1}[(\mathbf{b}_{j} \cdot \mathbf{b}_{j+1})/ (|\mathbf{b}_{j}| |\mathbf{b}_{j+1}|)]$ in terms of the bond vector $\mathbf{b}_{j} = \mathbf{r}_{j} - \mathbf{r}_{j-1}$ between two neighboring beads $j$ and $j-1$. The parameter associated with the rest angle is fixed at $B = 1.5$. We utilize $A = 0 \varepsilon$ and $6 \varepsilon$ to study the effect of chain rigidity on polymer glass formation.

The number of beads in a single chain is $M = 20$ in our simulations, and the total bead number of the whole system is $N = 8000$ or $12000$ for $A = 0 \varepsilon$ or $6 \varepsilon$. We describe all quantities and results from MD simulations in standard reduced LJ units. Specifically, length, time, temperature, and pressure are, respectively, given in units of $\sigma$, $\tau$, $\varepsilon / k_{\mathrm{B}}$, and $\varepsilon / \sigma^3$, where $\tau = \sqrt{m_b \sigma^2 / \varepsilon}$ with $m_b$ being the bead mass. The reduced units may be roughly mapped to laboratory units, e.g., by taking the suggested choices of Baschnagel and coworkers,~\cite{Baschnagel_Chapter} i.e., $\sigma \approx 5 \times 10^{-10}$ m, $\varepsilon / k_{\mathrm{B}} \approx 450$ K, and $m_b \approx 60$ g/mol. A reduced time of $t = 1 \tau$ and a reduced pressure of $P = 1 \varepsilon / \sigma^3$ then correspond approximately to $1$ ps and $50$ MPa, respectively. As a reference, the entanglement length is about $M = 84$ in a similar coarse-grained linear polymer melt without bending constraints at a number density of $\rho = N/V = 0.85 \sigma^{-3}$ and a temperature of $T = 1.0 \varepsilon/k_{\mathrm{B}}$.~\cite{2016_JCP_145_141101} The chain length with $20$ beads is well below the entanglement length, but long enough to be regarded as a polymer. Since quantitative comparisons between the GET and simulation is difficult at present, we chose a reasonable chain length to check the predictions qualitatively so that we can obtain simulation results for polymer melts within a reasonable amount of computational time. For a detailed discussion of the relationship between the simulation model and real polymers, the reader is recommended to ref~\citenum{2020_Mac_53_1901}, where mapping relations between the reduced and laboratory units have been provided for a wide range of commodity polymers by tuning the chain rigidity of the simulation model such that the number of Kuhn segments within the volume of a Kuhn length cube for the simulation model matches that for the target polymer.

The Large-scale Atomic/Molecular Massively Parallel Simulator (LAMMPS) molecular dynamics package~\cite{LAMMPS_1995, LAMMPS_webpage} is utilized to perform MD simulations in three dimensions under periodic boundary conditions, and a time step of $\Delta t = 0.005 \tau$ is used to integrate the equations of motion. For each polymer system, we initially prepare an equilibrated melt system at constant $P$ and $T$. The temperature is chosen to be sufficiently high so that the polymer melt can be equilibrated properly within our time window. This step requires us to first run simulations in the $NPT$ ensemble, where the number density is determined from a production run of $10^5 \tau$ under the specified thermodynamic conditions after an equilibration of $10^5 \tau$. Following the $NPT$ simulations, the melt with the desired density is further equilibrated for a period of $2 \times 10^5 \tau$ in the $NVT$ ensemble, which is much longer than the longest relaxation time of the melt to ensure the proper equilibration of the polymer melt. The equilibrated melt is subsequently subjected to a cooling or heating process at a constant $P$ at a rate of $10^{-4} \varepsilon / (k_{\mathrm{B}} \tau)$, which enables us to obtain the density as a function of $T$. In our simulations, we focus on a $T$ regime well above the glass transition temperature $T_{\mathrm{g}}$ so that our results are not complicated by the nonequilibirum effects associated with the glass state. Equilibrium properties are calculated in the $NVT$ ensemble after the melt is further equilibrated for a period typically over $10$ to $100$ times longer than the segmental structural relaxation time $\tau_{\alpha}$ determined from the self-intermediate scattering function (see more details in Section~\ref{Sec_ScalingDynamic}).

\section{Results and Discussion}

\subsection{\label{Sec_Relation}Interrelation between Thermodynamic Properties}

We first focus on the basic thermodynamic properties predicted by the thermodynamic component of the GET, the LCT. To this end, we consider a melt of PP chains, where a single bending energy is adequate to describe the chain rigidity.~\cite{2021_Mac_54_3001, 2008_ACP_137_125} In our calculations, the chain length is $N_c = 8000$, as defined by the number of repeating units in a single chain, which is typical of high molecular masses. This chain length lies in a mass range where the segmental dynamics and thermodynamic properties considered remain nearly unchanged with varying $N_c$. Our calculations for variable $N_c$ also indicate that the main findings and conclusions remain the same for other $N_c$ regarding the interrelations between thermodynamic properties and other results discussed below. Notably, the GET allows for direct computations of the characteristic temperatures of polymer glass formation, $T_A$, $T_c$, and $T_{\mathrm{g}}$, whose positions are shown below in the $T$ dependence of the properties as reference.

\begin{figure*}[htb]
	\centering
	\includegraphics[angle=0,width=0.8\textwidth]{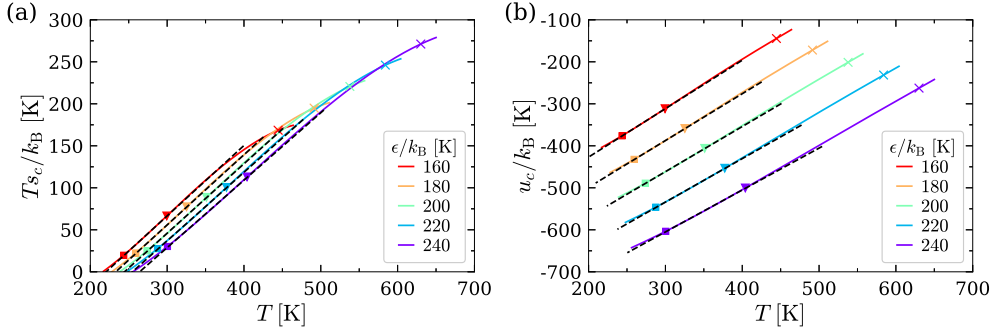}
	\caption{\label{Fig_GET_UkB}Temperature dependence of the thermodynamic properties of polymer glass formation predicted from the lattice cluster theory (LCT). Panels (a) and (b) correspond to the product of the temperature and configurational entropy density, $Ts_c / k_{\mathrm{B}}$, and the configurational internal energy density, $u_c / k_{\mathrm{B}}$, for varying $\epsilon$ at zero pressure for polymer melts with the PP structure having $N_c = 8000$ and $E_b/k_{\mathrm{B}} = 600$ K. The cross, triangle, and square symbols indicate the positions of $T_A$, $T_c$, and $T_{\mathrm{g}}$, respectively. Dashed lines are linear fits in the $T$ range between $T_c$ and $T_{\mathrm{g}}$.}
\end{figure*}

 Figure~\ref{Fig_GET_UkB} shows the $T$ dependence of $T s_c / k_{\mathrm{B}}$ and $u_c / k_{\mathrm{B}}$ for varying $\epsilon$ for the PP melt having $E_b/k_{\mathrm{B}} = 600$ K. Since the calculations are performed at zero pressure, the configurational enthalpy density is equal to the configurational internal energy density. The effect of pressure is discussed below. Concomitant with the dynamic slowing down upon cooling (Figure~\ref{Fig_GET_Tau}), we see that both $T s_c$ and $u_c$ decrease with decreasing $T$. In our previous works,~\cite{2021_Mac_54_3001, 2008_ACP_137_125} we have analyzed the $T$ dependence of $s_c$ in detail. In particular, over a limited $T$ range in the low-$T$ regime of glass formation between $T_c$ and $T_{\mathrm{g}}$, a linear relation between $Ts_c / k_{\mathrm{B}}$ and $T$ holds with the slope $K_T$,~\cite{2013_JCP_138_12A548}
 \begin{equation}
 	\label{Eq_TSc}
 	Ts_c / k_{\mathrm{B}} = K_T(T / T_K - 1),\ T_{\mathrm{g}} < T < T_c,
 \end{equation}
 where $T_K$ is the Kauzmann temperature at which $s_c$ is extrapolated to zero. Since $K_T$ bears no direct relation to the strength of the $T$ dependence of $\tau_{\alpha}$, we term this quantity the low $T$ fragility parameter. Equation~\ref{Eq_TSc} is shown as dashed lines in Figure~\ref{Fig_GET_UkB}a. We may thus obtain $T_K$ at which $T s_c$ extrapolates to zero in Figure~\ref{Fig_GET_UkB}a, although the applicability of the thermodynamic theory below $T_{\mathrm{g}}$ is questionable, where glass materials are normally out of equilibrium and the approximations upon which the LCT are based also lead to uncertainty. 
 
While extensive attention has been given to $s_c$ in the past due to its central role in the entropy theory, we have not analyzed the other fundamental thermodynamic quantities before, such as $u_c$ and $h_c$. Interestingly, Figure~\ref{Fig_GET_UkB}b indicates the linear dependence of $u_c$ and $h_c$ on $T$ over a wide range of $T$, as described by the following equations, 
 \begin{equation}
	\label{Eq_U}
	u_c / k_{\mathrm{B}} = K_{T,u}(T / T_K - 1) + u_{c,0} / k_{\mathrm{B}},
\end{equation}
and
 \begin{equation}
	\label{Eq_H}
	h_c / k_{\mathrm{B}} = K_{T,h}(T / T_K - 1) + h_{c,0} / k_{\mathrm{B}},
\end{equation}
where $K_{T,u}$ and $K_{T,h}$ are adjustable parameters and $u_{c,0}$ and $h_{c,0}$ are the extrapolated values of $u_{c}$ and $h_{c}$ at $T_K$. We checked that such a linear relationship of the $T$ variation of $u_{c}$ and $h_{c}$ exists for polymer melts having variable chain rigidity, chain length, pressure, and monomer structure. Hence, the above equations are general relationships predicted from the LCT. This analysis evidently points to a close connection between $Ts_c$, $u_{c}$, and $h_{c}$ as well as between the thermodynamics and dynamics, as shown below. Note that $u_{c,0}$ and $h_{c,0}$ can be uniquely defined by the values to which these properties extrapolate at $T_K$, which ensures compatibility with the VFT relation in models assuming that the reciprocals of these configurational properties determine the apparent activation energy in cooled liquids. Based on this ``consistency criterion'', we may then calculate $u_{c,0}$ and $h_{c,0}$ from the LCT.

 \begin{figure*}[htb]
	\centering
	\includegraphics[angle=0,width=0.8\textwidth]{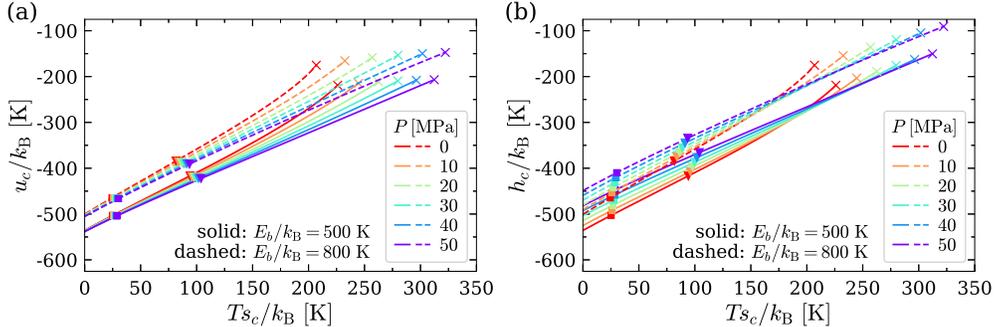}
	\caption{\label{Fig_GET_RelateTS}Correlation between the configurational internal energy density, the configurational enthalpy density, and the configurational entropy density predicted from the LCT. Panels (a) and (b) show $u_c / k_{\mathrm{B}}$ and $h_c / k_{\mathrm{B}}$ versus $Ts_c / k_{\mathrm{B}}$ for varying $P$ for polymer melts with the PP structure having $N_c = 8000$ and $\epsilon/k_{\mathrm{B}} = 200$ K. Solid and dashed lines correspond to the results for $E_b/k_{\mathrm{B}} = 500$ K and $800$ K, respectively. The cross, triangle, and square symbols indicate the positions of $T_A$, $T_c$, and $T_{\mathrm{g}}$, respectively.}
\end{figure*}

The above observation immediately implies a strong correlation between $T s_c$, $u_{c}$, and $h_{c}$ as $T$ is varied. To illustrate this behavior, Figure~\ref{Fig_GET_RelateTS} shows $u_{c} / k_{\mathrm{B}}$ and $h_{c} / k_{\mathrm{B}}$ versus $T s_c / k_{\mathrm{B}}$ for varying $P$. Here, the effect of chain rigidity is also considered. While curvature appears in the high-$T$ regime near $T_A$, a feature that is less pronounced at higher $P$, a linear relationship is evident between $T s_c$ and $u_{c}$ or $h_{c}$ in the $T$ range between $T_c$ and $T_{\mathrm{g}}$. Curvature above $T_c$ is expected, since it is well established that the VFT relation no longer applies, or at least a different VFT expression is often fitted empirically, in this high-$T$ regime of glass formation.~\cite{2015_PRE_92_062304} The GET offers a precise alternative to the VFT equation for quantifying relaxation dynamics in the high-$T$ regime of glass formation, but we do not dwell on this issue in the present paper, except to point out that the deviation from a linear scaling with $T$ at elevated $T$ is not necessarily a defect in the predicted relation between dynamics and thermodynamics.

The phenomenon in Figure~\ref{Fig_GET_RelateTS}, noted before from an experimental study of estimates of $u_c$ and $s_c$ by Caruthers and Medvedev,~\cite{2018_PRM_2_055604} is a particular example of entropy-enthalpy compensation, a widely occurring phenomenon in measurements and simulations of the thermodynamics of condensed fluids as well as in the dynamics of condensed fluids in connection to the enthalpy and entropy of thermal activation.~\cite{2001_CR_101_673, 2008_JPCB_112_15980, 1992_PRB_46_12244, 1986_JPC_19_5655, 1946_JCP_14_591, 1961_Nature_191_1292, 1943_TFS_39_48, 1959_PPS_73_153, 1940_JACS_62_3113, 2015_JCP_143_144905, 2014_NatCommun_5_4163, 2015_JCP_142_234907, 1995_PRE_51_1791, 2006_PRE_74_031501, 2008_JPCB_112_15980, 2012_SoftMatter_8_2983, 2013_SoftMatter_9_241} In summary, the LCT predicts that changes in the enthalpy and entropy in condensed systems tend to occur in a correlated fashion, a fact with numerous practical and scientific consequences.~\cite{2001_CR_101_673, 2008_JPCB_112_15980, 1994_JPC_98_1515, 2020_JCP_153_154901, 1976_JCP_65_4701, 1989_JAP_66_1308, 2017_EBJ_46_301, 1991_JMS_26_4477, 1989_SSC_69_707, 2009_PJ_41_455} The entropy-enthalpy compensation effect is particularly prevalent in the thermodynamics of molecular binding at surfaces and has been repeatedly shown in this context based on the LCT framework.~\cite{2015_JCP_142_214906, 2017_JCP_147_064909, 2019_JCP_151_124709} We will discuss this entropy-enthalpy compensation effect further below.

\subsection{\label{Sec_ScalingThermo}Thermodynamic Scaling of Thermodynamic Properties}

We can provide further insight into the relation between the thermodynamics and dynamics by examining the scaling behavior of these properties under different fixed $P$ conditions. In particular, numerous experimental and computational studies~\cite{2005_RPP_68_1405, 2010_Mac_43_7875, Book_Roland, Book_Paluch} have established that most liquids seem to exhibit a remarkable, yet poorly understood, property termed ``thermodynamic scaling'' in which the structural relaxation time $\tau_{\alpha}$, and many other dynamic properties, can be expressed in terms of a ``universal'' reduced variable, $TV^{\gamma}$, where $\gamma$ is a scaling exponent describing how $T$ and $V$ are linked to each other when either quantity is varied. In a previous work,~\cite{2021_Mac_54_3247} we showed that this scaling relation can be derived by combining the Murnaghan equation of state~\cite{1944_PNAS_30_244, Book_Murnaghan, 1995_IJT_16_1009} with the GET. Thermodynamic scaling arises in the non-Arrhenius relaxation regime as a scaling property of the fluid configurational entropy density $s_c$, normalized by its value $s_c^*$ at the onset temperature $T_A$ of glass formation, $s_c / s_c^*$, so that a constant value of $TV^{\gamma}$ corresponds to a \textit{reduced isoentropic} fluid condition.

\begin{figure*}[htb]
	\centering
	\includegraphics[angle=0,width=0.975\textwidth]{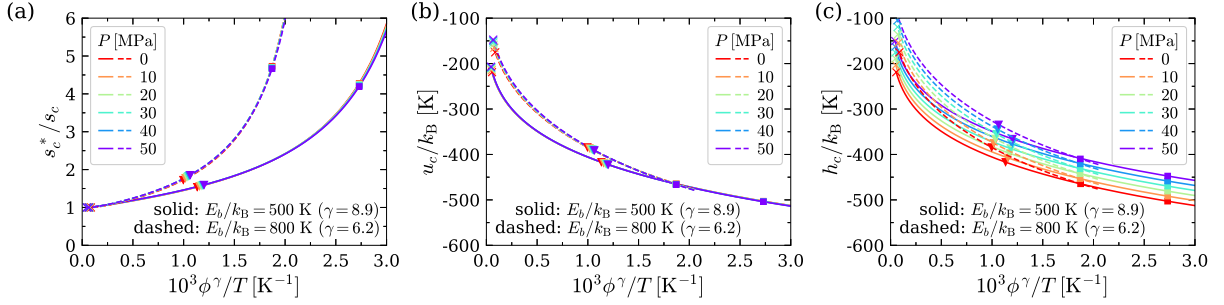}
	\caption{\label{Fig_GET_ScalingThermo}Test of thermodynamic scaling of the thermodynamic properties based on the LCT. Panels (a)--(c) show $s_c^*/s_c$, $u_c / k_{\mathrm{B}}$, and $h_c / k_{\mathrm{B}}$ versus $10^3 \phi^{\gamma}/T$ for varying $P$ for polymer melts with the PP structure having $N_c = 8000$ and $\epsilon/k_{\mathrm{B}} = 200$ K. Solid and dashed lines correspond to the results for $E_b/k_{\mathrm{B}} = 500$ K and $800$ K, respectively. The cross, triangle, and square symbols indicate the positions of $T_A$, $T_c$, and $T_{\mathrm{g}}$, respectively.}
\end{figure*}

\begin{figure*}[htb]
	\centering
	\includegraphics[angle=0,width=0.8\textwidth]{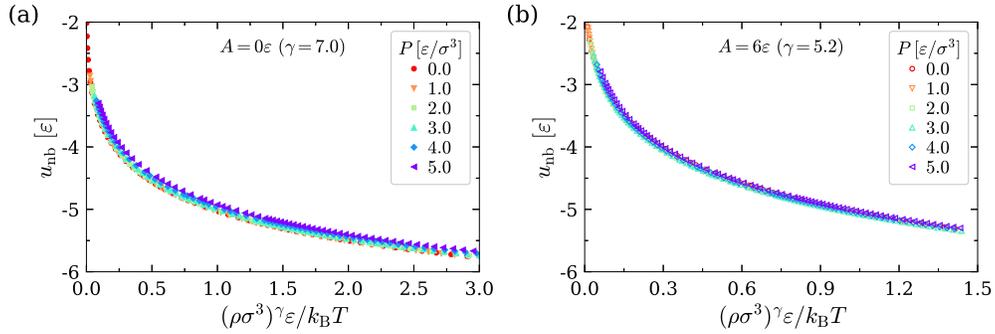}
	\caption{\label{Fig_MD_UPair}Test of thermodynamic scaling of the nonbonded potential energy based on the molecular dynamics (MD) simulations. Panels (a) and (b) show $u_{\mathrm{nb}}$ versus $(\rho \sigma^3)^{\gamma}\varepsilon/k_{\mathrm{B}}T$ for varying $P$ for $A = 0 \varepsilon$ and $6 \varepsilon$, respectively.}
\end{figure*}

Here, we extend our analysis to other thermodynamic properties, $u_{c}$ and $h_{c}$. We first show in Figure~\ref{Fig_GET_ScalingThermo}a that $s_c / s_c^*$ exhibits the scaling property with smaller $\gamma$ for stiffer chains, consistent with our previous study.~\cite{2013_JCP_138_234501} This trend is also consistent with our simulation results of a coarse-grained polymer model, as discussed below. It is evident from Figure~\ref{Fig_GET_ScalingThermo}b,c that $u_{c}$ obeys thermodynamic scaling, while $h_{c}$ does not. We suggest that this is due to the fact that the additional ``thermodynamic potential'' $PV$ contribution to $h_c$ is not consistent with thermodynamic scaling. 

Because the configurational thermodynamic properties are challenging to determine in simulations, here we consider a closely related quantity that is readily accessible from simulation, namely, the nonbonded potential energy, $u_{\mathrm{nb}} = U_{\mathrm{nb}} / N$. We see from Figure~\ref{Fig_MD_UPair} that $u_{\mathrm{nb}}$ obeys thermodynamic scaling in our coarse-grained polymer model, which seems to support the finding of the GET for the configurational internal energy density. Notably, Caruthers and Medvedev~\cite{2018_PRM_2_055604, 2019_Mac_52_1424} suggested that the configurational internal energy might exhibit thermodynamic scaling and that this possibility should be investigated. We next turn to the mysterious parameter $s_c^*$ of the AG and GET models, whose significance was noted before by Johari~\cite{2000_JCP_112_8958, 2000_JCP_112_7518} as a basic thermodynamic parameter of relevance to the dynamics of GF liquids, but which is normally treated as an adjustable constant in attempts at comparing the AG theory to experiment. We again see the advantage of the LCT, which allows for the direct computation of $s_c$ for a broad range of polymers under general thermodynamic conditions. We shall see that $s_c^*$ gives some important insights into the free energy surface governing the dynamics of polymeric GF liquids.

\begin{figure*}[htb]
	\centering
	\includegraphics[angle=0,width=0.8\textwidth]{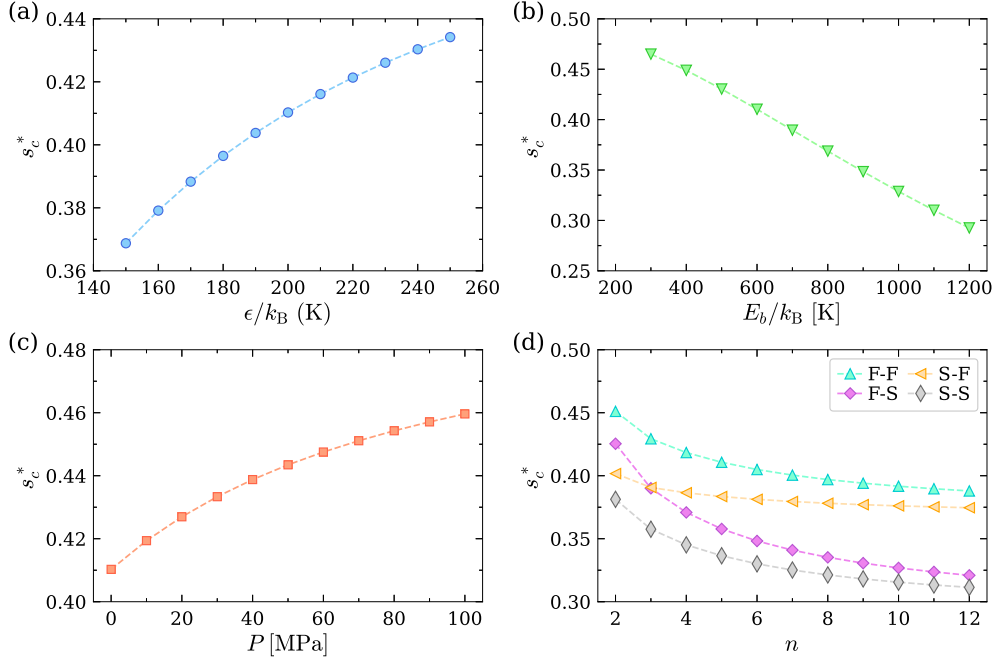}
	\caption{\label{Fig_GET_STA}Dependence of the high temperature limit of the configurational entropy density on molecular and thermodynamic parameters predicted from the LCT in the limit of high polymer mass. Panels (a)--(c) show $s_c^*$ as a function $\epsilon$, $E_b$, and $P$ for polymer melts with the PP structure, respectively. Panel (d) shows $s_c^*$ as a function of the side-chain length $n$ for different classes of polymers.}
\end{figure*}

\begin{figure*}[htb]
	\centering
	\includegraphics[angle=0,width=0.8\textwidth]{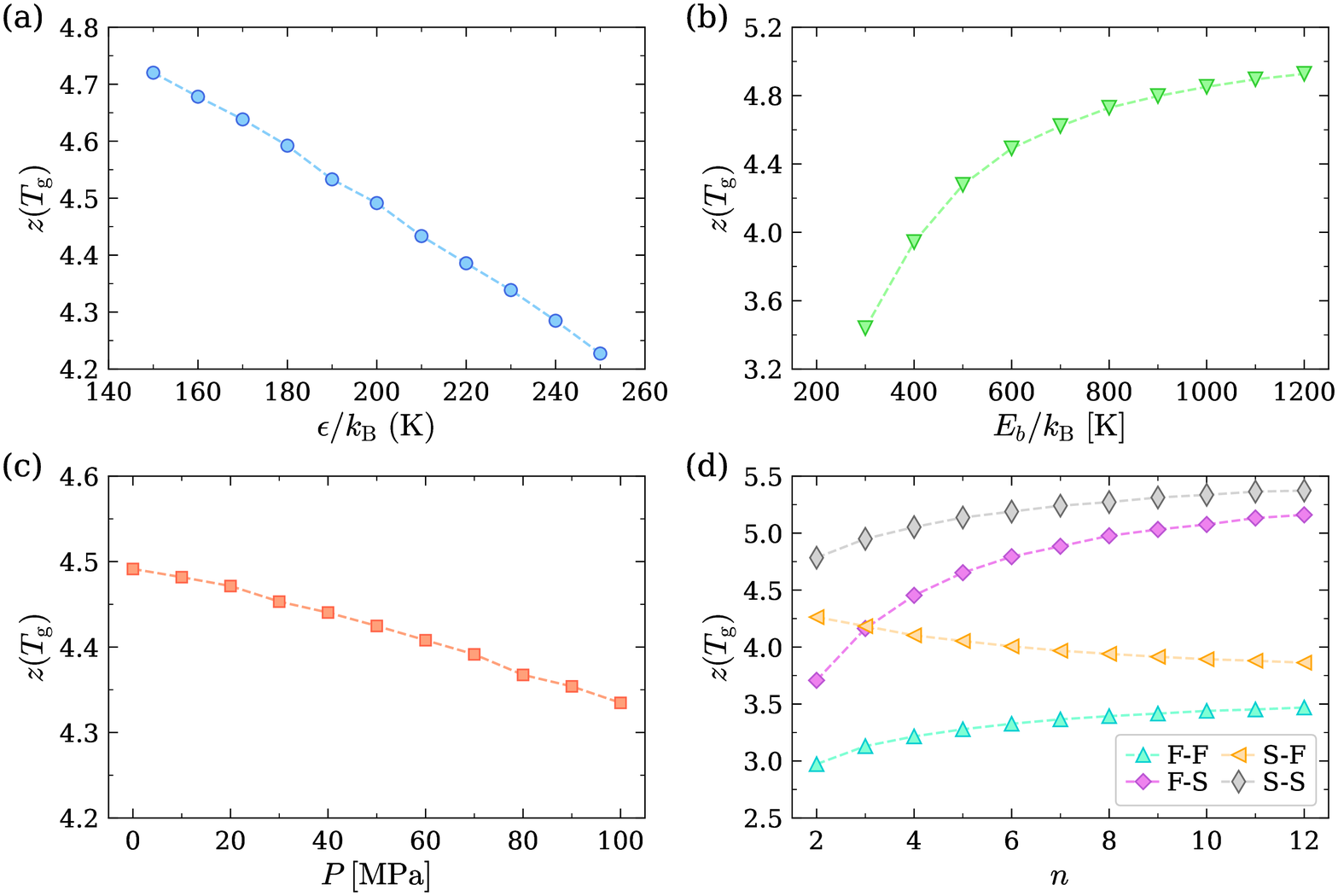}
	\caption{\label{Fig_GET_NSTg}Dependence of the extent of collective motion at the glass transition temperature on molecular and thermodynamic parameters predicted from the LCT in the limit of high polymer mass. Panels (a)--(c) show $z(T_{\mathrm{g}})$ as a function $\epsilon$, $E_b$, and $P$ for polymer melts with the PP structure, respectively. Panel (d) shows $z(T_{\mathrm{g}})$ as a function of the side-chain length $n$ for different classes of polymers.}
\end{figure*}

In the lattice model of polymer melts, the strength of the attractive interaction between the polymer segments is modulated by a square well-like potential where the depth of this potential is set by the well depth parameter, $\epsilon$, and its range is set by the lattice spacing. The effect of this cohesive interaction parameter on the associated free energy surface can be appreciated from the variation of $s_c^*$, the maximum in the configurational entropy density $s_c$ that occurs at an elevated $T$ where the material can freely access all the configurational minima and the dynamics is strongly stochastic. As noted in Section~\ref{Sec_Intro}, we may expect that increasing the strength of the attractive component of the pair potential should modulate the topography of the many-body free energy surface. This is the essential concept in the energy renormalization method of obtaining more realistic coarse-grained intermolecular potentials, which modulates the magnitude of $\epsilon$ as a function of $T$ to ``correct'' to the greatest extent inaccuracies in the $T$ dependence of $s_c(T)$ that arise from coarse-graining.~\cite{2019_SciAdv_5_eaav4683} We see from Figure~\ref{Fig_GET_STA}a that $s_c^*$ calculated from the LCT in the limit of long chains monotonically increases with increasing $\epsilon$. On the other hand, increasing the chain rigidity has the opposite effect of greatly reducing the number of minima in the energy surface, and hence, $s_c^*$ decreases as $E_b$ is increased (Figure~\ref{Fig_GET_STA}b), an obvious consequence of the reduction in the fluctuations in polymer shape. Perhaps less obviously, we find that increasing the applied pressure leads to an increase in $s_c^*$ (Figure~\ref{Fig_GET_STA}c), while increasing the side-chain length $n$ has the effect of reducing $s_c^*$ in different classes of polymer melts having the structure of poly($\alpha$-olefins). Here, we extend our analysis for polymer melts having the structure of PP to variable side-chain length $n$ and variable relative rigidities of the backbone and side chains, a difference that has served to define general classes of polymers.~\cite{2021_Mac_54_3001, 2008_ACP_137_125} In particular, we model different classes of polymers by choosing the bending energy parameters $E_b/k_{\mathrm{B}}$ and $E_s/k_{\mathrm{B}}$ for the backbone and side chains to be $200$ K and $200$ K for flexible polymers with relatively flexible side groups or F-F polymers, $200$ K and $600$ K for flexible polymers with relatively stiff side groups or F-S polymers, $600$ K and $200$ K for polymers having a relatively stiff backbone and flexible side groups or S-F polymers, and $600$ K and $600$ K for polymers having a relatively stiff backbone and relatively stiff side groups or S-S polymers, respectively. More generally, $s_c^*$ and other thermodynamic properties depend on the polymer mass, and the LCT allows for calculations of these corrections in a systematic fashion. As one might expect, these finite chain length effects track those of the characteristic temperatures of glass formation, including $T_{\mathrm{g}}$. We will report elsewhere on this type of finite chain length effect on the thermodynamics and dynamics of polymer liquids based on both the GET and MD simulations. As a related matter, we mention that $s_c^*$ scales with the spatial dimension $d$ as $s_c^* \sim \ln(d/2)$ in the limit of large $d$.~\cite{2016_ACP_161_443}

Varying molecular and thermodynamic parameters can also be expected to vary the ``roughness'' of the free energy surface, and this topographic attribute of the highly complex multidimensional surface can be quantified by the derivative of $s_c^*/ s_c(T)$ with respect to $T$, which is identified with the scale of collective motion in the AG~\cite{1965_JCP_43_139} and GET~\cite{2021_Mac_54_3001, 2008_ACP_137_125} models, $z(T) = s_c^*/ s_c(T)$. The differential variation of $z(T)$ provides a well-defined and sometimes observable ``dynamic fragility index'' that is not complicated by contributions of the high-$T$ activation energy as in the case of the commonly estimated fragility parameters, $m$, the differential activation energy at $T_{\mathrm{g}}$ (eq~\ref{Eq_Angell}), and the parameter $D$ defined by the VFT equation (eq~\ref{Eq_VFT}).~\cite{2020_Mac_53_7239} Future work is required to understand the relation between this dynamic fragility index and the fragility parameters of both the high- and low-$T$ regimes of glass formation, i.e., $C_s$ and $m$.

Finally, we note that the magnitude of $z(T)$ at the various characteristic temperatures can be directly calculated from the GET. This quantity is defined to equal unity at the onset temperature $T_A$ and increases monotonically towards a limiting value at $T_{\mathrm{g}}$, below which it becomes difficult to equilibrate the material. We show $z(T_{\mathrm{g}})$ as functions of $\epsilon$, $E_b$, $P$, and $n$ in Figure~\ref{Fig_GET_NSTg}. We see that $z(T_{\mathrm{g}})$ falls in a relatively narrow range between $\approx 3$ and $\approx 5.5$, so we tentatively judge that a ``typical'' value is $4$ for $z(T_{\mathrm{g}})$. A similar range for $z(T_{\mathrm{g}})$ has been found by Johari,~\cite{2000_JCP_112_8958} except for a couple of cases where $z(T_{\mathrm{g}})$ was estimated to be as large as $10$. Caruthers and Medvedev~\cite{2018_PRM_2_055604} deduced that the ratio of the activation energies at $T_A$ and $T_{\mathrm{g}}$ for $19$ materials lies in a range between $2$ and $7$. Moreover, the magnitude of $z(T_{\mathrm{g}})$ tends to be smaller in systems having a higher $s_c^*$. Evidently, the greater density of minima helps reduce the necessity of collective motion in the liquid at low $T$. This general behavior is apparent in the $T$ variation of $s_c$, as illustrated in ref~\citenum{2021_Mac_54_3001}, where we contrast the variation of a highly flexible polymer, characterized by a very strong type of glass formation, with that of a relatively stiff polymer exhibiting a relatively fragile glass formation. It is very helpful that the GET allows for the calculation of $s_c^*$ as a function of molecular and thermodynamic parameters, because this basic metrical parameter evidently encodes important information about the geometry of the free energy surface.

\subsection{\label{Sec_Transform}Transformation of the GET into Different ``Equivalent'' Thermodynamic Representations}

\begin{figure*}[htb]
	\centering
	\includegraphics[angle=0,width=0.975\textwidth]{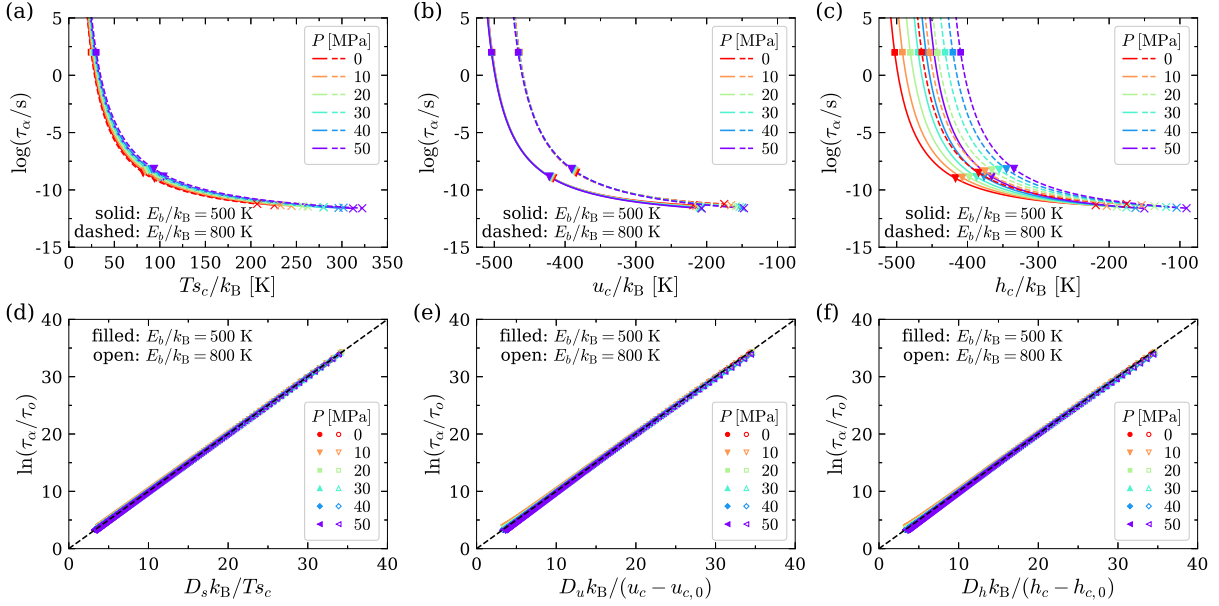}
	\caption{\label{Fig_GET_Model}Relationship between the thermodynamics and the dynamics predicted from the GET. Panels (a)--(c) show $\log \tau_{\alpha}$ versus $Ts_c / k_{\mathrm{B}}$, $u_c / k_{\mathrm{B}}$, and $h_c / k_{\mathrm{B}}$, respectively, for varying $P$ for polymer melts with the PP structure having $N_c = 8000$ and $\epsilon/k_{\mathrm{B}} = 200$ K. Solid and dashed lines correspond to the results for $E_b/k_{\mathrm{B}} = 500$ K and $800$ K, respectively. The cross, triangle, and square symbols indicate the positions of $T_A$, $T_c$, and $T_{\mathrm{g}}$, respectively. Panels (d)--(f) show the corresponding reduction based on eqs~\ref{Eq_VFTS}--\ref{Eq_VFTH}. Data are restricted to the $T$ range above $T_{\mathrm{g}}$. Filled and open symbols correspond to the results for $E_b/k_{\mathrm{B}} = 500$ K and $800$ K, respectively. Dashed lines indicate the equivalence of the two properties considered.}
\end{figure*}

Our discussion in Section~\ref{Sec_Relation} implies the presence of a direct relation between $\tau_{\alpha}$ and the configurational thermodynamic properties, $u_{c}$ or $h_{c}$, due to the interrelation between these properties and $Ts_c$. To demonstrate this explicitly, Figure~\ref{Fig_GET_Model}a--c shows $\log \tau_{\alpha}$ as a function of $T s_c / k_{\mathrm{B}}$, $u_{c} / k_{\mathrm{B}}$, and $h_{c} / k_{\mathrm{B}}$, respectively. Surprisingly, we see that the relation between $\log \tau_{\alpha}$ and $u_{c}$ is nearly independent of $P$. Further analysis indicates that the relationships between $\log \tau_{\alpha}$ and $T s_c$, $u_{c}$, and $h_{c}$ can be described by VFT-like equations,
\begin{equation}
	\label{Eq_VFTS}
	\tau_{\alpha} = \tau_o \exp \left(\frac{D_s k_{\mathrm{B}}}{Ts_c}\right),
\end{equation}
\begin{equation}
	\label{Eq_VFTU}
	\tau_{\alpha} = \tau_o \exp \left(\frac{D_u k_{\mathrm{B}}}{u_c - u_{c,0}}\right),
\end{equation}
and
\begin{equation}
	\label{Eq_VFTH}
	\tau_{\alpha} = \tau_o \exp \left(\frac{D_h k_{\mathrm{B}}}{h_c - h_{c,0}}\right),
\end{equation}
where $\tau_o = 10^{-13}$ s and $D_s = (\Delta G_o / k_{\mathrm{B}}) (s_c^* / k_{\mathrm{B}})$, as can be deduced from the AG relation in eq~\ref{Eq_AG}. The parameters $D_u$, $D_h$, $u_{c,0}$, and $h_{c,0}$ can be determined from eqs~\ref{Eq_TSc}--\ref{Eq_H}. The data reduction based on the above equations is presented in Figure~\ref{Fig_GET_Model}d--f. While eq~\ref{Eq_VFTS} is a direct result of the entropy theory, eqs~\ref{Eq_VFTU} and~\ref{Eq_VFTH} are not readily anticipated.

\begin{figure*}[htb]
	\centering
	\includegraphics[angle=0,width=0.975\textwidth]{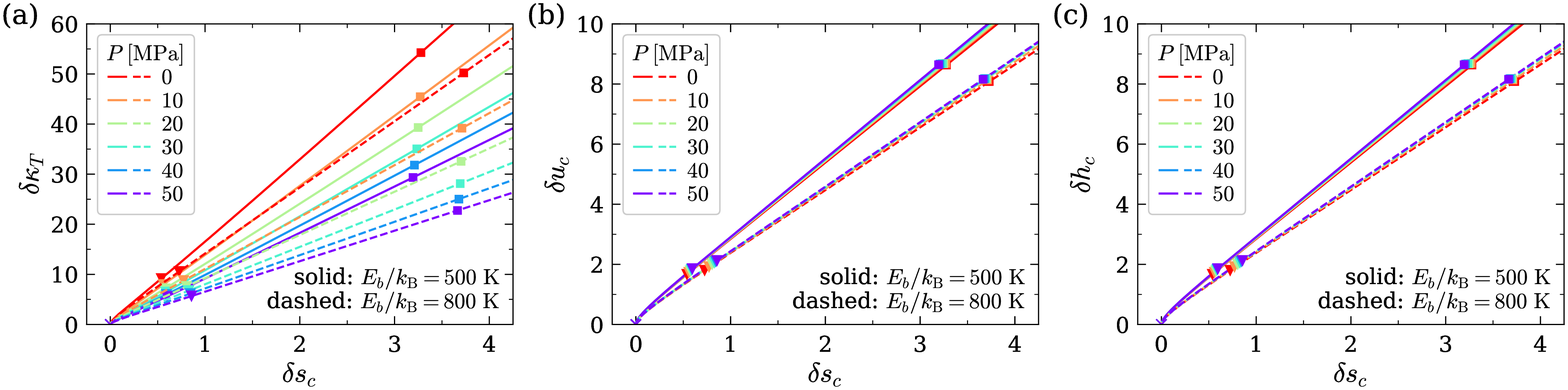}
	\caption{\label{Fig_GET_Transform}Relationship between the configurational entropy density and other thermodynamic properties predicted from the LCT. (a) Reduced isothermal compressibility $\delta \kappa_T$, (b) reduced configurational internal energy density $\delta u_c$, and (c) reduced configurational enthalpy density $\delta u_c$ versus reduced configurational entropy density $\delta s_c$ for varying $P$ for polymer melts with the PP structure having $N_c = 8000$ and $\epsilon/k_{\mathrm{B}} = 200$ K. Solid and dashed lines correspond to the results for $E_b/k_{\mathrm{B}} = 500$ K and $800$ K, respectively. The cross, triangle, and square symbols indicate the positions of $T_A$, $T_c$, and $T_{\mathrm{g}}$, respectively.}
\end{figure*}

The above result suggests deeper relations between the configurational thermodynamic properties. Motivated by our previous work,~\cite{2021_Mac_54_3247} we can define the following reduced quantities,
\begin{equation}
	\label{Eq_ScReduced}
	\delta s_c \equiv s_c^*/ s_c - 1 = (s_c^* - s_c)/s_c,
\end{equation} 
\begin{equation}
	\label{Eq_ICReduced}
	\delta \kappa_T \equiv (\kappa_{T,A} - \kappa_T) / (\kappa_T - \kappa_{T,o}),
\end{equation} 
\begin{equation}
	\label{Eq_UReduced}
	\delta u_c \equiv (u_{c,A} - u_c) / (u_c - u_{c,o}),
\end{equation}
and 
\begin{equation}
	\label{Eq_HReduced}
	\delta h_c \equiv (h_{c,A} - h_c) / (h_c - h_{c,o}).
\end{equation} 
where $\kappa_{T,A}$, $u_{c,A}$, and $h_{c,A}$ are the corresponding values of $\kappa_{T}$, $u_{c}$, and $h_{c}$ at $T_A$ and $\kappa_{T,o}$, $u_{c,o}$, and $h_{c,o}$ are the corresponding values of $\kappa_{T}$, $u_{c}$, and $h_{c}$ at the temperature at which $s_c$ extrapolates to $0$ as determined directly from the LCT. This leads to a proportional relation between $\delta s_c$ and $\delta \kappa_T$, $\delta u_c$, or $\delta h_c$, as shown in Figure~\ref{Fig_GET_Transform}. Based on this transformation, it is readily shown that the reduced quantities normalized by the proportional factor exhibit thermodynamic scaling, since $\delta s_c$ exhibits thermodynamic scaling. It should be noted that while we have some latitude in which thermodynamic property is used to relate to $\tau_{\alpha}$, $s_c$ retains a particular significance because the reference temperature is defined by the limit at which $s_c(T)$ extrapolates to zero. In this sense, $s_c$ is a ``special'' thermodynamic property with regard to glass formation.

\subsection{\label{Sec_ScalingDynamic}Thermodynamic Scaling of Segmental Relaxation Time, Extent of Cooperative Motion, and ``Slow'' $\beta$-Relaxation}

We may extend the line of reasoning discussed above to develop theories of glass formation based on other thermodynamic properties. It is well known that the density $\rho$ of GF materials often exhibits a linear $T$ dependence over a wide $T$ range above $T_{\mathrm{g}}$,~\cite{1967_PRSL_298_481} as for the other thermodynamic properties, so that the relaxation time can be modeled by the classical ``free volume'' expression, 
\begin{equation}
	\label{Eq_VFTRho}
	\tau_{\alpha} = \tau_o \exp \left(\frac{D_{\rho} \rho}{\rho_0 - \rho}\right),
\end{equation} 
where $D_{\rho}$ and $\rho_0$ are phenomenological parameters. We previously showed that the same argument leads to an expression that fits simulation observations very well.~\cite{2016_Mac_49_8341, 2016_Mac_49_8355} The same type of reasoning can be extended to other thermodynamic properties such as the isothermal compressibility, where the thermodynamic property serves as a \textit{surrogate} for temperature. Of course, temperature has the advantage of relative simplicity, and we have chosen to represent all our results in terms of $T$ to avoid the conceptual and technical problems associated with experimental estimates of $s_c$. 

We may gain some insight into which thermodynamic properties might bear a more fundamental relation to dynamics, by checking which properties exhibit the property of thermodynamic scaling of the form exhibited by dynamical properties in relation to structural relaxation. In a previous work,~\cite{2021_Mac_54_3247} we found that the isothermal compressibility does not obey this symmetry based on the GET and simulations. This lack of scaling is also found for the configurational enthalpy in our discussion above based on the GET. Hence, it is evident that some thermodynamic properties conform to thermodynamic scaling, while others do not. The lack of thermodynamic scaling of the isothermal compressibility $\kappa_T$ implies that this type of scaling unfortunately does not arise from the intermolecular potential having a power-law form, which is the popular rationalization of this scaling.~\cite{2021_Mac_54_3247} It is notable in this connection that the thermodynamic scaling exponent $\gamma$ is highly variable in the GET model, and in our simulation estimates of certain thermodynamic properties, even though the pair potential is \textit{fixed} by a form similar to an off-lattice square-well potential.~\cite{2021_Mac_54_3247, 2013_JCP_138_234501} We must then look for another origin of thermodynamic scaling in both the dynamics and thermodynamics of molecular liquids. At present, the origin of this apparently ``universal'' scaling property is frankly obscure.

\begin{figure*}[htb]
	\centering
	\includegraphics[angle=0,width=0.8\textwidth]{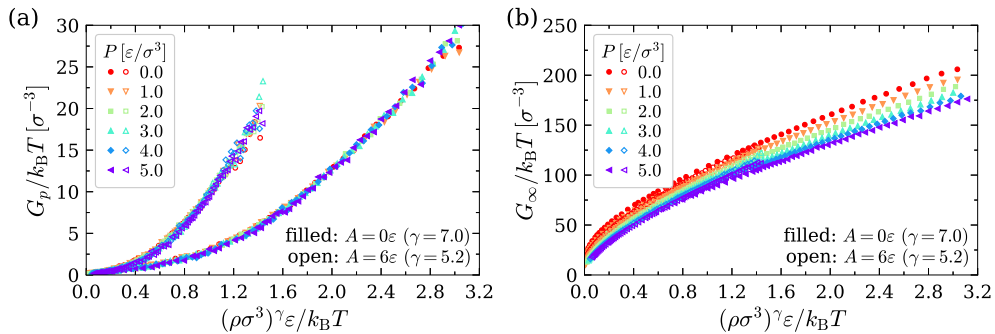}
	\caption{\label{Fig_MD_GT}Test of thermodynamic scaling of the shear modulus based on the MD simulations. Panels (a) and (b) show the glassy plateau and instantaneous shear modulus normalized by $k_{\mathrm{B}} T$, corresponding to $G_{p} / k_{\mathrm{B}} T$ and $G_{\infty} / k_{\mathrm{B}} T$, versus $(\rho \sigma^3)^{\gamma}\varepsilon/k_{\mathrm{B}} T$ for varying $P$ for $A = 0 \varepsilon$ and $6 \varepsilon$, respectively.}
\end{figure*}

The GET indicates that thermodynamic scaling has other subtle aspects that offer potential clues to the origin of this phenomenon. For example, $s_c(T)$ does not by itself exhibit this scaling, while the ratio $s_c^*/ s_c (T)$ does.~\cite{2021_Mac_54_3247} Further, Leporini and coworkers~\cite{2012_JCP_136_041104, 2016_JCP_145_234904, 2017_JPCM_29_135101} have emphasized an aspect of this scaling relation that might be crucially important for its occurrence. They observed that this scaling arises in the glassy plateau $G_p$ in the shear stress relaxation function divided by $k_{\mathrm{B}} T$ and in the mean-square displacement on a ``caging'' timescale on the order of a picosecond (i.e., the ``Debye-Waller parameter'' $\langle u^2 \rangle$),~\cite{2016_JCP_145_234904, 2017_JPCM_29_135101} and they further emphasized earlier arguments by Tobolsky~\cite{Book_Tobolsky} that $G_p$ was largely predominated by intermolecular interactions so that this property is related to the cohesive energy density of the liquid. Atomic motions not involving bond displacements clearly dominate the magnitude of $\langle u^2 \rangle$, which Leporini and coworkers also found to be directly related to $G_p$ in a universal way.~\cite{2016_JCP_145_234904, 2017_JPCM_29_135101} We have confirmed these results in our own coarse-grained simulations of polymer fluids in Figure~\ref{Fig_MD_GT}a. Following the works of Leporini and coworkers,~\cite{2016_JCP_145_234904, 2017_JPCM_29_135101} the glassy plateau $G_p$ is determined from the stress autocorrelation function,
\begin{equation}
	\label{Eq_GT}
	G(t) = \frac{V}{k_{\mathrm{B}} T} \left< \sigma_{xy}(t)\sigma_{xy}(0) \right>
\end{equation}
where $\sigma_{xy}$ is the off-diagonal component of the stress tensor and the brackets $\left< \ldots \right>$ denote the usual thermal average. We have utilized the multiple-tau correlator method of Ram\'{i}rez et al.~\cite{2010_JCP_133_154103} for the calculation of $G(t)$. We then obtain the glassy plateau as $G(t)$ evaluated at the caging time $t = 1 \tau$, $G_p = G(t = 1 \tau)$, and the infinite frequency shear modulus is determined from the limit, $G_{\infty} = G(t = 0)$.

The observation in Figure~\ref{Fig_MD_GT}a leads us to consider how these observations might pertain to common thermodynamic properties, such as the configurational enthalpy and internal energy. A direct computation of the nonbonded potential energy $u_{\mathrm{nb}}$, where the enthalpic contributions from the chemical bonds are not included by construction, indicates that this property indeed obeys thermodynamic scaling (Figure~\ref{Fig_MD_UPair}). Moreover, thermodynamic scaling is also found for the configurational internal energy calculated from the GET, as described above. We thus hypothesize that thermodynamic scaling only arises in molecular fluid properties of both a thermodynamic and dynamic nature in which bonded interactions are not prevalent. Consistent with this working hypothesis, direct computations from simulations of the bulk modulus $B$ and the infinite frequency shear modulus $G_{\infty}$, properties for which the bond interactions are clearly important, indicate that these properties do not follow thermodynamic scaling, as shown in our previous work~\cite{2021_Mac_54_3247} for $B$ and in Figure~\ref{Fig_MD_GT}b for $G_{\infty} / k_{\mathrm{B}} T$, respectively. This possible rationale for the origin of thermodynamic scaling in terms of intermolecular interactions deserves further investigation, but we view this approach as promising.

It should be noted that it is sometimes possible to subject properties to linear transformations to obtain reduced properties that exhibit thermodynamic scaling even if the initial thermodynamic variable itself does not have this property, as indicated in Section~\ref{Sec_Transform}. The absence of thermodynamic scaling by a given thermodynamic property, e.g., isothermal compressibility, does not by itself generally preclude the use of that thermodynamic property in describing the relaxation time or other dynamic property. For example, our previous work~\cite{2021_Mac_54_3247} has shown based on the GET that the isothermal compressibility, which certainly does not exhibit thermodynamic scaling, can be subjected to a linear transformation relating this property to a reduced form of $s_c$, a quantity that exhibits thermodynamic scaling to a high degree of approximation. The ultimate test for the suitability of a theoretical model of the relaxation time or other dynamic property of interest must then be whether or not the predicted expression obeys thermodynamic scaling.

\begin{figure*}[htb]
	\centering
	\includegraphics[angle=0,width=0.8\textwidth]{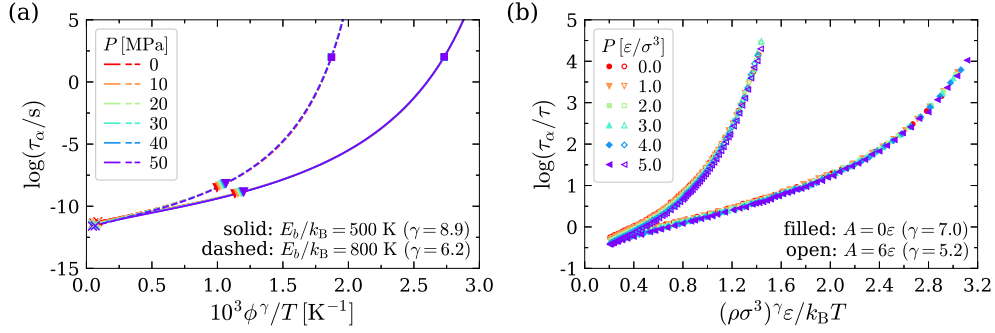}
	\caption{\label{Fig_ScalingTau}Test of thermodynamic scaling of the structural relaxation time. (a) $\log \tau_{\alpha}$ versus $10^3 \phi^{\gamma}/T$ calculated from the GET for varying $P$ for polymer melts with the PP structure having $N_c = 8000$ and $\epsilon/k_{\mathrm{B}} = 200$ K. Solid and dashed lines correspond to the results for $E_b/k_{\mathrm{B}} = 500$ K and $800$ K, respectively. The cross, triangle, and square symbols indicate the positions of $T_A$, $T_c$, and $T_{\mathrm{g}}$, respectively. (b) $\log \tau_{\alpha}$ versus $(\rho \sigma^3)^{\gamma}\varepsilon/k_{\mathrm{B}}T$ obtained from the MD simulations for varying $P$. Filled and open symbols correspond to the results for $A = 0 \varepsilon$ and $6 \varepsilon$, respectively.}
\end{figure*}

\begin{figure*}[htb]
	\centering
	\includegraphics[angle=0,width=0.8\textwidth]{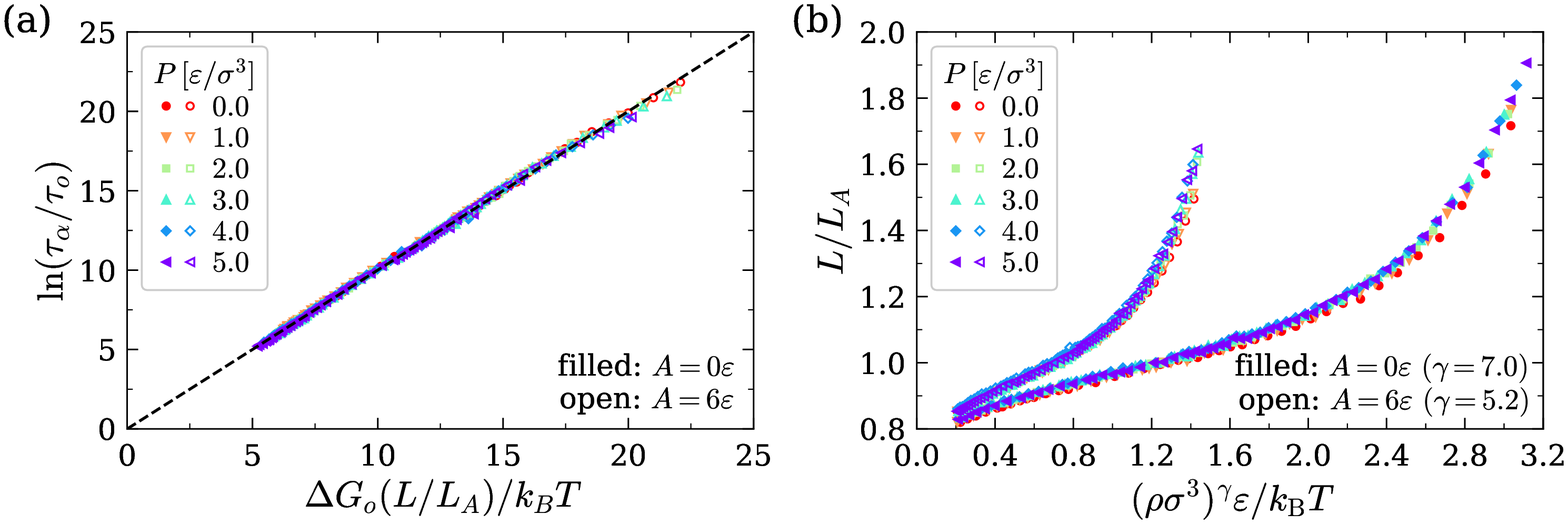}
	\caption{\label{Fig_String}Test of the string model of glass formation and thermodynamic scaling of the extent of cooperative motion based on the MD simulations. (a) String model description of the relationship between the structural relaxation time and the extent of cooperative motion. The dashed lines indicate $\ln (\tau_{\alpha}/\tau_o) = \Delta G_o(L/L_A)/k_{\mathrm{B}}T$, where the determinations of $\tau_o$, $\Delta G_o$, and $T_A$ follow the method described elsewhere.~\cite{2020_Mac_53_4796, 2020_Mac_53_6828, 2020_Mac_53_9678, 2021_Mac_54_9587, 2022_Mac_55_3221} (b) $L / L_A$ versus $(\rho \sigma^3)^{\gamma}\varepsilon/k_{\mathrm{B}}T$ for varying $P$. Filled and open symbols correspond to the results for $A = 0 \varepsilon$ and $6 \varepsilon$, respectively.}
\end{figure*}

Since our interest in thermodynamic scaling derives from the apparent general occurrence of this scaling in experimental systems, we next consider a simulation ``reality check'' by examining the thermodynamic scaling of the segmental relaxation time $\tau_{\alpha}$ for our coarse-grained model. As in our previous works,~\citenum{2020_Mac_53_4796, 2020_Mac_53_6828, 2020_Mac_53_9678, 2021_Mac_54_9587, 2022_Mac_55_3221} $\tau_{\alpha}$ is defined by the time at which the self-intermediate scattering function $F_s(q, t)$ decays to $0.2$, where the wavenumber is chosen to be $q = 7.0 \sigma^{-1}$ corresponding approximately to the first peak position of the static structure factor. Both our calculations based on the GET and simulation results show that the $T$ and $\rho$ dependences of $\tau_{\alpha}$ indeed conform to thermodynamic scaling in our model polymer melts having different chain rigidities for a range of $P$, as shown in Figure~\ref{Fig_ScalingTau}, where the scaling index $\gamma$ is altered by chain rigidity. These results confirm those found in our previous paper~\cite{2021_Mac_54_3247} reviewing the thermodynamic scaling phenomenon, and thus, are not surprising.

We are now in a position to test the consistency of the string model of the dynamics of GF liquids, a close conceptual relative of the GET model, with thermodynamic scaling. In Figure~\ref{Fig_String}a, we show the fit of the same $\tau_{\alpha}$ data to the string model of glass formation, following exactly the same method described elsewhere,~\cite{2020_Mac_53_4796, 2020_Mac_53_6828, 2020_Mac_53_9678, 2021_Mac_54_9587, 2022_Mac_55_3221} where the accord is found to be excellent over the $T$ range accessible to our simulations. This analysis yields as one of its outputs, which can be calculated independently of $\tau_{\alpha}$,~\cite{2014_JCP_140_204509, 2021_Mac_54_3001} an estimate of the scale of particle exchange collective motion $L$, relative to its value at the onset temperature $T_A$, i.e., $L(T_A) \equiv L_A$. The ratio, $L / L_A$, is the analog of the size of the CRR, $z = s_c^* / s_c(T)$, of the GET model~\cite{2021_Mac_54_3001} and should by consistency exhibit thermodynamic scaling. We see in Figure~\ref{Fig_String}b that $L / L_A$ indeed exhibits thermodynamic scaling. A previous simulation study of a similar coarse-grained polymer model~\cite{2013_JCP_138_12A541} confirmed the relation, $L \sim 1 / S_c$, to a good approximation, a basic premise on which the AG and GET models are based if $z$ is identified with $L / L_A$. It is also worth mentioning that, since the lowest temperatures accessible to our simulations are close to $T_c$, $L / L_A$ at $T_c$ falls roughly in a range between $1.5$ and $1.9$ in our simulations (Figure~\ref{Fig_String}b), which is in line with the predictions of the GET (Figure~\ref{Fig_GET_ScalingThermo}a).

\begin{figure*}[htb]
	\centering
	\includegraphics[angle=0,width=0.8\textwidth]{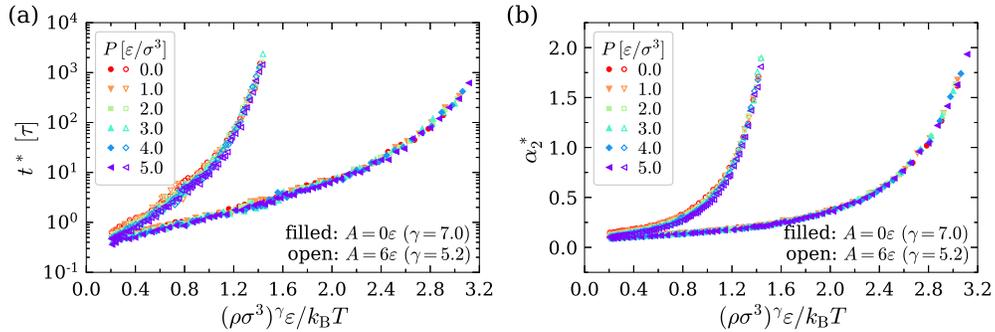}
	\caption{\label{Fig_ScalingNGP}Test of thermodynamic scaling of the non-Gaussian parameter based on the MD simulations. Panels (a) and (b) show the peak time $t^*$ and magnitude $\alpha_2^*$ of the non-Gaussian parameter versus $(\rho \sigma^3)^{\gamma}\varepsilon / k_{\mathrm{B}}T$ for varying $P$. Filled and open symbols correspond to the results for $A = 0 \varepsilon$ and $6 \varepsilon$, respectively.}
\end{figure*}

The GET is currently limited to making statements about the segmental relaxation time in polymeric GF liquids, and there are evidently other dynamical properties that can be measured, and which are of considerable interest in applications. In particular, recent studies have made the hypothesis that the JG $\beta$-relaxation time $\tau_{\mathrm{JG}}$,~\cite{1970_JCP_53_2372, 1971_JCP_55_4245, 1973_JCP_58_1766} the relaxation process of primary importance in the glass state, can be identified with the peak time $t^*$ in the non-Gaussian parameter, $\alpha_2(t)$.~\cite{2021_JCP_154_084505, 2021_EPJE_44_56} This proposed relationship, which is still currently a hypothesis supported by simulation observations, has many testable implications, given the fact that intensive attention for both time scales has been given in previous studies of GF liquids. Because $t^*$ has been observed computationally to exhibit a general ``decoupling'' scaling relationship,~\cite{2013_JCP_138_12A541, 2016_Mac_49_8355, 2017_Mac_50_2585, 2020_Mac_53_4796} i.e., $(t^* / \tau_{f}) \sim (\tau_{\alpha} / \tau_{f})^{1 - \zeta}$, where $\tau_{f}$ is the ``fast'' $\beta$-relaxation time on the order of a picosecond, as in the case of the $\tau_{\mathrm{JG}}$ experimentally~\cite{2005_Mac_38_7033} and computationally,~\cite{2021_JCP_154_084505, 2021_EPJE_44_56} we may infer that $t^*$ should also obey thermodynamic scaling, as confirmed in Figure~\ref{Fig_ScalingNGP}a, where the thermodynamic scaling exponent $\gamma$ is exactly the same as shown above for $\tau_{\alpha}$. We have previously shown that the ``decoupling exponent'' $\zeta$ and $\tau_{f}$ in our coarse-grained model are \textit{highly insensitive} to pressure,~\cite{2017_Mac_50_2585} so that $t^*$ is predicted to be \textit{invariant} when $P$ and $T$ are varied in such a way that $\tau_{\alpha}$ is fixed in magnitude. This invariance relation is a commonly reported phenomenological feature of $\tau_{\mathrm{JG}}$ in small-molecule liquids, where it has also been shown that both $\tau_{\mathrm{JG}}$ and $\tau_{\alpha}$ exhibit thermodynamic scaling,~\cite{2007_JPCM_19_205133, 2012_JCP_137_034511, 2015_JNCS_407_98} as we have found for $t^*$ and $\tau_{\alpha}$. We note, however, that this ``invariance relation'' between $\tau_{\mathrm{JG}}$ and $\tau_{\alpha}$ no longer holds if the material is perturbed in such a way that the exponent $\zeta$ is altered, e.g., when molecular additives or nanoparticles are added to the GF material, thereby altering molecular packing and intermolecular cohesion so that the fragility of the GF liquid, and thus $\zeta$, changes.~\cite{2008_JCP_128_034709, 2016_JSM_5_054048} In this situation, the magnitude of $\tau_{\mathrm{JG}}$ and $\tau_{\alpha}$ often shift in opposite directions with the additive. Additives that reduce packing frustration and thus fragility have the effect of increasing $\tau_{\mathrm{JG}}$ in the $T$ regime below $T_{\mathrm{g}}$ where this relaxation is normally measured, but $\tau_{\alpha}$ is decreased. Such additives are ``antiplasticizers''.~\cite{2010_SoftMatter_6_292, 2006_PRL_97_045502, 2007_JCP_126_234903} Nanoparticle additives are particularly interesting in that their addition to polymer matrices can increase both fragility and $\zeta$,~\cite{2011_PRL_106_115702} which leads to smaller $\tau_{\mathrm{JG}}$ and larger $\tau_{\alpha}$.~\cite{2015_Polymer_68_47, 2010_JMS_153_79, 2009_Mac_42_3201} The nanoparticle additives can thus achieve the reverse of ``antiplasticization'', which must be distinguished from ``plasticized'' materials in which the additive does not alter $\zeta$ significantly so that both $\tau_{\alpha}$ and $\tau_{\mathrm{JG}}$ shift in the same direction by the additive, as in the case of the pressurization measurements mentioned above. There are a huge range of applications revolving around these fragility modifying additives.

We may also make some general statements about the activation energy of the JG $\beta$-relaxation process based on the decoupling relation between this relaxation time and $\tau_{\alpha}$ and the activation energy of $\tau_{\alpha}$ in relation to $T_c$ or $T_{\mathrm{g}}$ discussed above, along with the magnitude of the scale of the collective motion, $z(T_{\mathrm{g}})$. If we take the ``typical'' estimates of $\Delta H_o \approx 10 k_{\mathrm{B}} T_{\mathrm{g}}$, $z \approx 4$, and $\zeta \approx 2 / 5$,~\cite{1998_JNCS_235_237_137} then we estimate the activation energy of the JG $\beta$-relaxation process to be approximately $\Delta H_{\mathrm{JG}} \approx 24 k_{\mathrm{B}} T_{\mathrm{g}}$. This estimate is very much in line with observations where the same prefactor has been estimated,~\cite{1997_EPL_40_649} although the prefactor is found to significantly fluctuate around this average value when many GF systems are considered.~\cite{2004_PRE_69_031501}

The phenomenological relation implies that the string length also governs the $T$ dependence of $t^*$ and $\tau_{\mathrm{JG}}$. Many of our results then carry over to describe the ``slow'' $\beta$-relaxation time. This time normally scales with the lifetime of the mobile particle clusters,~\cite{2013_JCP_138_12A541, 2015_JCP_142_164506} while $\tau_{\alpha}$ scales with the average lifetime of the immobile particle clusters.~\cite{2013_JCP_138_12A541, 2015_JCP_142_164506} By extension, the relaxation time $t_{\chi}$ associated with the $4$-point susceptibility $\chi_4$, which also scales with the lifetime of the immobile particle clusters,~\cite{2013_JCP_138_12A541, 2015_JCP_142_164506} also obeys thermodynamic scaling. We checked that it is indeed the case in our simulations (data not shown).

Finally, we consider one of the most commonly studied, but perhaps least understood measures of dynamic heterogeneity in GF liquids, $\alpha_2(t)$. While the peak time $t^*$ has been found to scale with the lifetime of mobile particles in both coarse-grained polymeric~\cite{2013_JCP_138_12A541} and Zr-Cu metallic GF materials,~\cite{2015_JCP_142_164506} and its thermodynamic scaling noted above, the peak magnitude, $\alpha_2^* \equiv \alpha_2(t^*)$, has often been reported and interpreted abstractly as some kind of measure of the intensity of the dynamic heterogeneity in cooled liquids. Zhang and Douglas~\cite{2013_SoftMatter_9_1254} have argued that $\alpha_2^*$ provides a measure of mobility fluctuations through its relation to a $4$-point velocity autocorrelation function, but this proposed relation requires further investigation. While $\alpha_2^*$ is simply zero by construction for Brownian particles, Odagaki and Hiwatari~\cite{1991_PRA_43_1103} have gained some insight into $\alpha_2^*$ by noting that this quantity is not zero for particles exhibiting jumping motions to a finite distance in the liquid where both theoretical and simulation studies of this type of diffusion process leads to the relation, $\alpha_2^* \ t^* = \mathrm{constant}$. The existence of a relation of this kind or a similar relation implies that $\alpha_2^*$ should also exhibit thermodynamic scaling, and we confirm in Figure~\ref{Fig_ScalingNGP}b that this scaling holds to a high degree of approximation, even though the simple scaling $\alpha_2^* \ t^*$ breaks down below $T_A$ where the fluid becomes dynamically heterogeneous and particle motion can no longer be described as simple diffusion. Curiously, we observe the specific approximate scaling relation at $T < T_A$, $\ln \alpha_2^* \sim 1/ \langle u^2 \rangle \sim G_p / k_{\mathrm{B}}T$, whose specific physical meaning we do not currently understand. Thermodynamic scaling provides a guiding theoretical thread that we continue to follow.

It is apparent from the discussion above that the thermodynamic scaling linking dynamic properties and specific thermodynamic properties exhibiting this scaling symmetry offers a powerful tool to understand the fundamental nature of models of GF liquids and the interrelation between relaxation processes of interest because they preserve this scaling property. As a final example of the thermodynamic-dynamic linkage of relevance to the dynamics of GF liquids, we make more comments regarding the JG $\beta$-relaxation process.~\cite{1970_JCP_53_2372, 1971_JCP_55_4245, 1973_JCP_58_1766} Some authors have recently suggested that the relaxation in structural glasses, which is known to be dominated by the JG $\beta$-relaxation process, reflects a physical situation in which the dynamics of the material is dominated by a hierarchical energy landscape having an ultrametric structure, i.e., the potential energy are organized in nested levels like a Cayley tree in their topology, as identified earlier in spin glasses near this glass transition.~\cite{1985_PRL_55_1634, 1984_PRL_53_958} The recent prediction of a Gardner transition in hard-sphere GF liquids at sufficiently high density is consistent with this point of view,~\cite{2018_PRL_120_085705, 2019_JCP_151_010901, 2020_PRL_124_078002} but it has not yet been established that this Gardner phase exists in molecular fluids in three dimensions.~\cite{2019_JCP_151_010901, 2017_PRL_119_205501} This ultrametric structure of the energy landscape in glasses was identified long ago in pioneering work on spin glasses, where the energy landscape was also determined to have an ultrametric structure,~\cite{1984_PRL_52_1156} and soon thereafter, this led to modeling of relaxation in glasses in terms of hopping processes in model ultrametric spaces.~\cite{1985_PRL_55_1634} Models of this kind arrived at some important conclusions of relaxation occurring in such energy landscapes. Blumen and coworkers~\cite{1986_JPA_19_L77, 1986_JPA_24_2807} and Vainas~\cite{1988_JPC_21_L341, 1991_JPCM_3_3941} have modeled the dynamics of systems constrained to ultrametric spaces and having random-walk type, where the relaxation time between different levels exactly obeys entropy-enthalpy compensation between the activation energy for ``hopping'' on the landscape and the entropy term in this model is governed by the branching index from one level of the energy landscape to another. This model also exhibits a nontrivial ergodic to ergodic transition at a characteristic temperature determined by the branching index of the hierarchical energy landscape at which the particle displacement changes from being transient to recurrent with an associated power-law relaxation at long times.~\cite{2021_JCP_154_084505, 1999_JPCM_11_A329, Douglas_Book} This type of dynamical transition is central to the theory of glass formation of Odagaki.~\cite{1995_PRL_75_3701} Intermittency of this kind is signaled by mobility and energy fluctuations having the form of colored noise.~\cite{2021_JCP_154_084505, 2013_SoftMatter_9_1254, 2013_SoftMatter_9_1266} This type of long-time power-law relaxation is characteristic of the JG $\beta$-relaxation process, along with a prevalence for the entropy-enthalpy compensation between the activation energy and entropy of this relaxation process.~\cite{2006_PRE_74_031501, 2012_SoftMatter_8_2983, 2008_JPCB_112_15980} Doliwa and Heuer~\cite{2003_PRE_67_031506} obtained significant insight into the energy landscape origin of the entropy-enthalpy compensation by studying the escape time from initial metabasin states having prescribed potential energies where these initial states were identified from an inherent structure analysis of the potential energy surface. The activation energy for the escape from the metabasin increased with the magnitude of the metabasin energy, while the logarithm of the prefactor of the Arrhenius escape rate varied linearly with the activation energy governing the escape rate, corresponding to the classic entropy-enthalpy compensation.

We note that the universality of the organization of the energy landscape into well-defined tiers of energy associated with an ultrametric structure is not clear, however. The matter of the topological structure of the free energy landscape associated with real molecular GF liquids is an outstanding question that deserves a careful investigation, although the general concept that the dynamics at low $T$ should be dominated by the topological structure of the free energy landscape seems to provide a reasonable qualitative perspective for understanding the dynamics in this low-$T$ ``glass'' regime. An interesting aspect of the model of the JG $\beta$-relaxation in terms of diffusion on an ultrametric energy surface mentioned above is that the long-time power exponent, describing the power-law decay of relaxation on this structure arising from hopping diffusively on the nodes of this surface corresponding to energy minima, depends linearly on temperature. On the other hand, in the alternative plausible scenario in which the energy surface is taken to have a self-similar hierarchical in the form of fractal structure,~\cite{2014_NatCommun_5_3725} such as a geometrical percolation cluster near the percolation threshold, a temperature independent power-law decay is obtained, as illustrated in the explicit calculations of Fujiawara and Yonezawa.~\cite{1995_PRL_74_4229} An important property of percolation clusters in large dimensions, i.e., $d > 6$, is that the spectral dimension governing the probability of return to a point (state) on such a surface is exactly $4/3$.~\cite{2009_IM_178_635} This fact, which is even a reasonable approximation for $d<6$ and for many other network structures,~\cite{1982_JPL_43_625} is important since the energy surface exists in a high dimensional space and the spectral dimension is normally highly dependent on spatial dimension. For example, the spectral dimension of an ordinary Euclidean lattice, such the cubic lattice, equals the spatial dimension in which it is embedded. It would appear from these models that we may learn essential information about the topological structure of the energy surface, such as its essential topological structure, from the temperature dependence of the long-time power law of the JG $\beta$-relaxation and the decay of the intermediate scattering function at low temperatures. While a linear temperature dependence of the power-law exponent describing the long-time decay of the $\beta$-relaxation has been observed in GF liquids,~\cite{1991_PRL_66_1334, 1998_PRB_58_14888} this exponent has also been shown to be nearly constant in others.~\cite{2000_JCP_112_2319, 1998_PRB_58_14888, 2021_JCP_154_084505} From the discussion above, there is some variability in the qualitative topological structure of the complex energy landscapes of GF liquids. We discuss further evidence supporting the existence of a hierarchical energy landscape for some simulated liquids in the next section.

\subsection{\label{Sec_Implication}Implications of Thermodynamic-Dynamic Interrelations}

If the dynamics of materials is related to the thermodynamic properties, then one might naturally expect such a relation to be inherent in the metrical properties of the energy landscape, beyond a consideration of the number of accessible minima in these surfaces as $T$ or other thermodynamic variables are changed. This brings us back to the energy landscape description of liquids discussed in the introduction (Section~\ref{Sec_Intro}) and ongoing work to render this qualitative picture quantitative. Most simulations have focused on the calculation of the configurational entropy of liquids by sampling the minima through a series of quenches to find the energy minima, as discussed in Section~\ref{Sec_Intro}, and this procedure is usually performed at constant volume so that the method quantifies a potential energy surface rather than a free energy surface.

The methodology mentioned above provides no information about the energies of the saddle points of the energy surface, but it is currently possible to determine the ``inherent structure'' energies of the energy basins in the potential energy surface so that we can begin to develop quantitative metrologies of the landscape itself. In particular, it is possible to designate the energies of the potential minima and connections representing abstract connecting paths on this complex energy surface, a construct termed the ``disconnectivity diagram''. Wales and coworkers~\cite{Book_Wales, 1998_Nature_394_758} and others~\cite{2002_JCP_117_10894, 2006_PNAS_103_18551} have shown the value of this type of coarse-grained roadmap of the topology of the energy surface for understanding the dynamics of many-body systems, especially in the case of nanoparticles and small biological molecules where this type of analysis can be made rather thoroughly, and some analyses of GF liquids of relatively small system size have been performed based on this methodology.~\cite{2001_PRB_64_024205, 1997_PD_107_225, 2003_JCP_118_3891, 2004_PNAS_101_14766} Yip and coworkers~\cite{2009_JCP_130_224504, 2009_JPCM_21_504104} have introduced a clever method for constructing disconnectivity diagrams for relatively large systems based on MD simulations, time series analysis of potential fluctuations, and a ``basin-filling algorithm'' constructed so that calculations cover the whole potential energy surface. This type of analysis quantitatively confirms the hierarchical nature of the energy landscape organized into well-defined ``tiers'' in which the minima have a similar energy with connections between these tiers having a tree-like structure reminiscent of the model of the energy surface by a Cayley tree, a model hyperbolic structure that can be represented readily on a lattice to enable analytic calculations. This type of hierarchical structure has been studied for many years in the context of the dynamics of proteins,~\cite{2001_PNAS_98_2370, 2003_PNAS_100_15534} but this type of potential energy structure is apparently rather common in the dynamics of condensed materials. This type of structure then allows for the study of metrical properties of the energy surface, such as the energy spacing between the tiers and branching index describing the connections between the tiers that are not specific to any given material system. This type of analysis is ongoing and promises to be very fruitful in illuminating the relationship between the energetics of these energy surfaces and the dynamics.

While no doubt useful, the highly schematic representation of the energy surface of many-body systems does not do justice to the high dimensional nature of these surfaces. This type of representation also does not address the complex structure and variable dimensionality of the saddle point regions through which connections between the energy minima occur. In the fluid regime, the trajectories in phase space describing the evolution of the material system spend most of their time in these saddle regions, so we must also be concerned with the energetics of these regions and the unstable collective modes associated with these unstable critical points of the energy surface.~\cite{1989_JCP_91_5581, 1991_JCP_94_6762} In a fluctuating ``random surfaces'' of this kind, one naturally expects the saddle points to be geometrically nested between the energy minima~\cite{1995_PRE_52_2348} so that these structures should likewise exhibit a hierarchy of ``quantized'' levels. Indeed, recent computational studies of model GF liquids aimed at quantifying the energies of the saddle points by energy minimization methods have suggested such a general structure that establishes a clear link to thermodynamic properties. In particular, the estimated average energy of the saddles was found by a number of groups for a range of model GF liquids~\cite{2000_PRL_85_5356, 2003_JCP_119_2120, 2004_JCP_121_7533, 2000_PRL_85_5360} to increase linearly with the order $k$ of the saddle where energy is measured with respect to the energy minimum, inherent structure energy, to which the saddles are related. Interestingly, anelastic measurements on metallic glass materials over a large frequency range have revealed direct evidence for a ``quantized'' activation energy spectrum for the secondary $\beta$-relaxation in these materials.~\cite{2014_AM_74_183, 2019_PRE_100_033001}

The energy landscapes found numerically for various simulated GF liquids indeed exhibit a remarkably general structure, resembling superficially the well-known energy level spacing characteristic of quantum harmonic oscillator. It seems also relevant to point out here that many-body dynamical systems have been observed to exhibit a classical analog of zero-point energy~\cite{1972_PLA_38_403, 1972_PRL_28_1173, 1980_PRA_22_1709, 1974_PRA_9_1252} associated with the emergence of an energy threshold, where the dynamic of the material first exhibits ergodic or ``stochastic'' dynamics rather than integrable or ``regular'' motion.~\cite{2013_PRA_2_2013, 1996_PRE_54_964} The interesting implications of this type of dynamical transition to foundational problems in physics have recently been reviewed by Carati et al.~\cite{2005_Chaos_15_015105}

Importantly, for the purposes of the present paper, the level spacing $\Delta E$ between these tiers of the energy surface by this metric was found to be $\Delta E \approx 10 k_{\mathrm{B}} T_c$, suggesting a direct link between the energy landscape structure and the thermodynamics of the fluid, a point which we amplify on below. We thus get the first hints of how the dynamics of liquids might be encoded in the metrical properties of their energy surfaces. It must be admitted at this stage that the energy minimization calculations required to determine the saddle points are much more difficult than the determination of the energy minima. Doye and Wales~\cite{2002_JCP_116_3777, 2003_JCP_119_12409} have shown that many of the sampled ``saddles'' in the simulation studies, just mentioned, were actually inflection points instead. Angelani et al.~\cite{2002_JCP_116_10297} and others~\cite{2002_PRL_88_255501} have ``qualified'' the earlier analysis by calling these characteristic points of the energy surface ``quasi-saddles'', where they further argue that the qualitative conclusions made in their earlier work should remain intact, regardless of this numerical difficulty. Wales and Doye~\cite{2003_JCP_119_12409} have shown how to overcome this numerical difficulty in finding the saddle points uniquely by a general analysis of regularities in the energetic values of the saddle points, but this method has so far not been attempted to reproduce the findings of Angelani et al.~\cite{2002_JCP_116_10297}

At this point, it is worth pointing out the potential importance of the higher-order saddles in relation to the emergence of collective particle motion in the form of strings. Keyes and coworkers~\cite{1993_JCP_98_3342} have argued that dimensionality of the saddles or their ``order'' bears a direct relationship to the emergence of collective motion in the material. In particular, reduction of the saddle degree should cause a focusing of the motion, akin to water flowing through a canyon, so that number of particles moving collectively should be proportional to reduction of the saddle degree, an argument which, if validated, would provide a possible energy landscape explanation of the emergence of string-like collective motion and the essential tenet of the AG theory that such collective motion should dominate the dynamics of liquids at low temperatures, where such collective exchange events become necessary to reach the higher energetic regions on the surface required to achieve large changes in configurational structure. 

Parisi and coworkers~\cite{2002_PRL_88_055502} adapted the interesting, if somewhat technically problematic, analysis of Angelani et al.~\cite{2002_JCP_116_10297} and Broderix et al.,~\cite{2000_PRL_85_5360} 
to a higher level by repeating the analysis of these authors and using this framework to tentatively estimate the barrier height between energy minima in the potential energy minima as a function of the energy density of the fluid, a quantity directly related to $T$. Encouragingly, their analysis led to qualitatively similar results for the temperature dependence of the average activation barrier height to the independent analysis by Yip and coworkers~\cite{2009_JCP_130_224504, 2009_JPCM_21_504104} mentioned above. This work seems to indicate that the energy barrier height varies with energy density, and thus temperature, well above $T_c$, and thus $T$ (see Figure 1 in ref~\citenum{2002_PRL_88_055502} for the specific relation between energy density and $T$), but it saturates to a constant value at much higher $T$ whose value is less than half the barrier height at $T_c$. The simulations also suggest that the barrier height saturates to a constant value at low $T$, as Yip and coworkers~\cite{2009_JCP_130_224504, 2009_JPCM_21_504104} have concluded. Overall, the barrier height for the model GF liquid was found with $T$ by a factor of about $4$. Angelani et al.~\cite{2002_JCP_116_10297} also inferred that the activation energy $\Delta H$ for diffusion estimated near $T_c$ was generally close to be $\Delta H(T_c) \approx 2 \Delta E$ for a range of soft-sphere liquids. Since the activation energy at $T_c$ has been found experimentally to be normally twice its value at $T_A$,~\cite{1999_JPCB_103_5895, 2006_JCP_125_144907} this correlation implies the suggestive relation, $\Delta H_o \approx \Delta E$. The GET also indicates that the activation energy at $T_c$ is generally about twice its high temperature value,~\cite{2021_Mac_54_3001, 2008_ACP_137_125} $\Delta H_o$. Numerous studies have established that $\Delta H_o$ can be estimated by thermodynamic properties,~\cite{2021_JCP_155_174901, 2015_JCP_142_234907, 2015_JCP_143_144905} so this provides the necessary link between thermodynamics and dynamics within the GET model. In the GET, $\Delta H_o$ is for simplicity approximated by $\Delta H_o \approx 6 k_{\mathrm{B}} T_c$, which is based on the empirical observation for a range of fluids and is consistent with the estimate of Angelini et al.~\cite{2002_JCP_116_10297} based on the landscape analysis. Once $\Delta H_o$ is prescribed by thermodynamic information in the GET, the theory can make predictions of the structural relaxation time based on purely thermodynamic information.

Importantly, the type of analysis above not only emphasizes the existence of activated transport above $T_c$, which is something denied in some models of liquid relaxation,~\cite{1998_Nature_393_554} but it also indicates a specific energy landscape interpretation of this quantity. This picture of activated transport and the origin of collective motion in many-body systems clearly deserve further investigation.

\section{Summary}

While the search for a fully predictive and theoretical satisfying model of the dynamics of glass-forming liquids that explains how the thermodynamics of these materials relate to their dynamics is still illusive, we have strived in the present work to at least understand the interrelation between thermodynamic properties proposed to be of interest in relation to the dynamics of liquids in various proposed models of the dynamics based on a general thermodynamic framework, the lattice cluster theory. After revealing a close interrelation between the configurational entropy, enthalpy, and internal energy, we have shown that the generalized entropy theory of glass formation can be recast into ``equivalent'' model expressions for the segmental relaxation time expressed in terms of alternative thermodynamic properties. While the configurational entropy has some theoretical aspects that favor this thermodynamic property over others, the difficulty of its experimental determination is definitely a drawback of this property so that other more readily measured properties provide useful practical alternatives in applications. 

The property of ``thermodynamic scaling'' has been found to be a powerful tool for discriminating which thermodynamic properties are most correlated with changes in the dynamics, which in turn raises the question of why some thermodynamic properties follow this scaling while others do not. The investigation of this scaling has led us to propose that the thermodynamic properties of molecular fluids dominated by intermolecular interactions rather than bond interactions exhibit this scaling property, a working hypothesis that provides some direction in developing new relationships interrelating dynamical properties as well as thermodynamic properties in the same ``class''. As another important result, the combination of the lattice cluster theory, generalized entropy theory, and molecular dynamics simulation has allowed us to leverage our understanding of the structural relaxation time in relation to thermodynamic properties to offer insights into dynamical properties such as the Johari-Goldstein $\beta$-relaxation process. Altogether, our analysis is broadly consistent with the hypothesis of the existence of deep interrelations between the dynamics and thermodynamics of glass-forming liquids, a thread that we continue to follow in new directions.

%% classes. This should be given within the "acknowledgement"
\begin{acknowledgement}
W.-S.X. acknowledges the support from the National Natural Science Foundation of China (Nos. 22222307 and 21973089). X.X. acknowledges the support from the National Natural Science Foundation of China (Nos. 21873092 and 21790341). W.-S.X. gratefully acknowledges HZWTECH for providing computation facilities. W.-S.X. and X.X. are grateful to Dr. Teng Lu for help with numerical calculations conducted on SunRising-1 computing environment, where a portion of the simulations in this work was done. This research also used resources of the Network and Computing Center at Changchun Institute of Applied Chemistry, Chinese Academy of Sciences.
\end{acknowledgement}

%\begin{suppinfo}
% Description of the Supporting Information.
%\end{suppinfo}

\bibliography{refs}

\end{document}